\documentclass[12pt]{article}

\usepackage[utf8]{inputenc}
\usepackage{latexsym, graphicx} 
\usepackage{amsmath,amsfonts,amssymb}
\usepackage{slashed}
\usepackage{cancel}
\usepackage{mathrsfs}
\usepackage{braket}
\usepackage{tensor}
\usepackage{xcolor}
\usepackage{cite}
\usepackage[colorlinks=true,linkcolor=blue,citecolor=blue,urlcolor=blue,filecolor=black]{hyperref}
\usepackage{tikz}
\usetikzlibrary{decorations.pathmorphing, calc}

\usepackage{pict2e}
\makeatletter
\newcommand{\loplus}{\mathbin{\mathpalette\dog@lsemi{+}}}
\newcommand{\dog@lsemi}[2]{\dog@semi{#1}{#2}{270,90}}
\newcommand{\dog@semi}[3]{%
	\begingroup
	\sbox\z@{$\m@th#1#2$}%
	\setlength{\unitlength}{\dimexpr\ht\z@+\dp\z@\relax}%
	\makebox[\wd\z@]{\raisebox{-\dp\z@}{%
			\begin{picture}(1,1)
			\linethickness{\variable@rule{#1}}
			\roundcap
			\put(0.5,0.5){\makebox(0,0){\raisebox{\dp\z@}{$\m@th#1#2$}}}
			\put(0.5,0.5){\arc[#3]{0.5}}
			\end{picture}%
	}}%
	\endgroup
}
\newcommand{\variable@rule}[1]{%
	\fontdimen8  
	\ifx#1\displaystyle\textfont3\else
	\ifx#1\textstyle\textfont3\else
	\ifx#1\scriptstyle\scriptfont3\else
	\scriptscriptfont3\relax
	\fi\fi\fi
}
\makeatother

\renewenvironment{thebibliography}[1]{
	\begin{oldthebibliography}{#1}
		\setlength{\itemsep}{2pt}
		\setlength{\parskip}{2pt}
	}
	{
	\end{oldthebibliography}
}

\numberwithin{equation}{section} 

\usepackage[listings]{tcolorbox}
\tcbuselibrary{listings,theorems}

\newtcbtheorem[number within=section]{mytheo}{Exercise}%
{colback=white,colframe=blue!30,fonttitle=\bfseries}{th}

\newtcbtheorem{program}{Bootstrap Program}%
{colback=white,colframe=blue!30,fonttitle=\bfseries}{th}

\newcommand{\mM}{\mathcal{M}}
\newcommand{\tmM}{\tilde{\mathcal{M}}}
\newcommand{\tg}{\tilde{g}}

\newcommand*{\hateq}{\mathop{}\! \hat{=} \mathop{}\! }

\newcommand*{\scri}{\ensuremath{\mathscr{I}}}
\newcommand*{\lied}{\mathop{}\!\mathcal{L}}
\newcommand*{\dd}{\mathop{}\!d}
\newcommand*{\sgn}{\text{sgn}}
\newcommand*{\zbar}{\bar{z}}

\newcommand*{\zb}{\bar{z}}
\newcommand*{\hb}{\bar{h}}

\newcommand*{\x}{\textbf{x}}

\begin{document}
	
	\begin{titlepage}
		\thispagestyle{empty}
		
		\begin{flushright}
		\end{flushright}
		
		\vskip1cm
		
		\begin{center}  
			{\LARGE\textbf{Lectures on Carrollian Holography}}
			
			\vskip1cm
			
			\centerline{\large Kevin Nguyen}
			
			\vskip1cm
			
			{\it{Universit\'e Libre de Bruxelles and Inernational Solvay Institutes,\\ ULB-Campus Plaine
					CP231, 1050 Brussels, Belgium}}\\
			\vskip 1cm
			{kevin.nguyen2@ulb.be}
			
		\end{center}
		
		\vskip1cm
		
		\begin{abstract} 
			Carrollian Holography aims to provide a holographic description of quantum gravity in asymptotically flat spacetimes, in terms of a novel kind of `carrollian' conformal field theory defined on the spacetime null conformal boundary $\scri$.
			
			The goal of these lectures is to propose a self-contained, pedagogical introduction to this active field of research. The main focus is given to the correspondence between massless scattering amplitudes, including gravitational amplitudes, and correlators in carrollian conformal field theory. Strong emphasis is put on the development of carrollian conformal field theory as an intrinsic, independent, and non-perturbative framework in which to formulate gravitational scattering theory.  
		\end{abstract}
		
		\vskip 5cm
		
		\begin{center}
			\emph{\large Based on Lectures given at Université Libre de Bruxelles during the Doctoral Solvay School,  October 2025.}
		\end{center}

	\end{titlepage}
	
	{\hypersetup{linkcolor=black}
		\tableofcontents 
	}
	
	
	\section{Asymptotically flat spacetimes}
	\label{sec: AFS}
	
	Carrollian Holography finds its roots in the study of \textit{asymptotically flat spacetimes}, a large class of solutions of General Relativity with zero cosmological constant $\Lambda=0$ used to model gravitational physics below the Hubble scale. Hence, we start our journey with a brief review of this classical subject, before heading for a wonderland where gravity meets the quantum.    
	
	\paragraph{Conventions.} A manifold $\tmM$ equipped with a lorentzian metric $\tg_{\mu \nu}$ is called a \textit{spacetime}. The hatted equality sign $\hateq$ refers to an equality at null infinity $\scri$. Indices $\mu,\nu,\rho,...$ denote four-dimensional, $\alpha,\beta,\gamma,...$ denote three-dimensional, and $i,j,k,..$ denote two-dimensional coordinate indices, respectively.
	
	\subsection{Bondi--Sachs description}
	\label{subsection: Bondi Sachs}
\begin{figure}[h!]
		\centering
		\includegraphics[clip,scale=0.5]{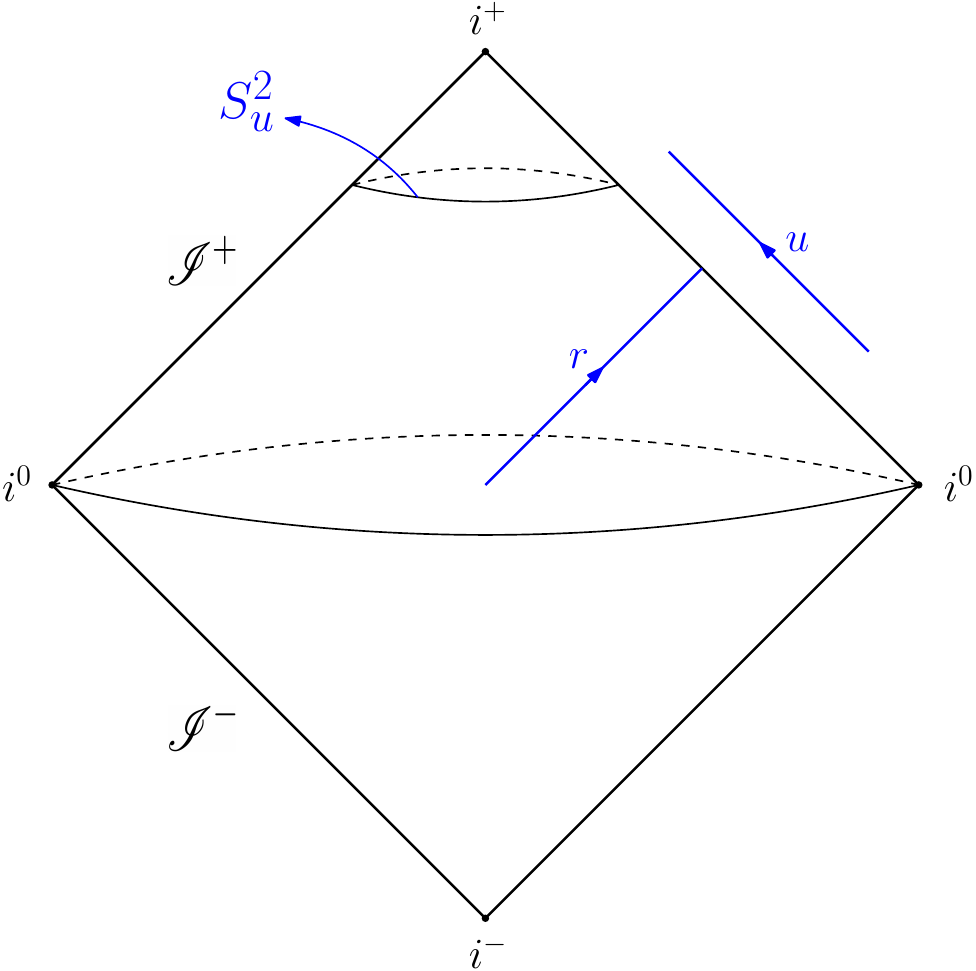}
		\caption{Penrose diagram of Minkowski spacetime $\mathbb{M}^4$, using retarded coordinates $x^\mu=(r,u,\vec x)$. Future null infinity $\scri^+$ is the constant-$r$ surface in the limit $r\to \infty$. The celestial sphere $\mathbb{S}^2_u$, covered by coordinates $\vec x$, is a `cut' of $\scri^+$ at constant $u$.}
		\label{Penrose}
	\end{figure}

	We want to study gravity coupled to matter, restricting to the case of zero cosmological constant $\Lambda=0$ and four spacetime dimensions. Obviously the vacuum solution of this theory is classically described by flat Minkowski spacetime $\mathbb{M}^4$. Choosing radial coordinates, its metric takes the simple form
	\begin{equation}
	\dd \tilde s^2=-\dd t^2+\dd r^2+r^2\, q_{ij} \dd x^i \dd x^j\,,
	\end{equation}
	where $q_{ij}$ is the unit round sphere metric in arbitrary coordinates $x^i$. For obvious physical reasons, we call this unit sphere $\mathbb{S}^2$ the \textit{celestial sphere} (also known as \textit{sky}). If one in interested in propagation of light, or any other massless field, it is useful to use retarded time $u=t-r$, such that 
	\begin{equation}
	\label{Minkowski in Bondi}
	\dd \tilde s^2=-\dd u^2-2\dd u \dd r+r^2\, q_{ij} \dd x^i \dd x^j\,.
	\end{equation}
	Lines of constant $(u,x^i)$ are null rays as easily seen from this metric, with $r$ the corresponding affine parameter. See Figure~\ref{Penrose}.
	
	While spacetime is not flat in presence of matter, there is certainly a sense in which it is approximately flat away from matter sources. More precisely, we will call a spacetime \textit{asymptotically flat} if there exists a coordinate system $(r,u,x^i)$ such that the metric can be written 
	\begin{equation}
	\label{Bondi gauge}
	\dd\tilde s^2=g_{uu} \dd u^2+2\, g_{ur} \dd u \dd r+2\, g_{ui} \dd u \dd x^i+ g_{ij} \dd x^i \dd x^j\,,
	\end{equation}
	and its components admit the large-$r$ expansion 
	\begin{equation}
	\label{Bondi expansion}
	\begin{split}
	g_{uu}&=-\frac{R_q}{2}+\frac{2M}{r}+o(r^{-1})\,,\\
	g_{ur}&=-1+\frac{1}{16r^2}\, C_{ij} C^{ij}+o(r^{-2})\,,\\
	g_{ui}&=\frac{1}{2} D^j C_{ij}+\frac{2}{3r}\, (N_i+\frac{1}{4} C_{ij} D_k C^{jk} )+ o(r^{-1})\,,\\
	g_{ij}&=r^2\, q_{ij}+r\, C_{ij}+\frac{1}{4} C_{mn} C^{mn}\, q_{ij}+o(r^0)\,.
	\end{split}
	\end{equation}
	Here $q_{ij}$ is the metric on the unit round sphere $\mathbb{S}^2$, or a metric in the same conformal class, $D_i$ the corresponding Levi-Civita connection, and $R_q$ the corresponding curvature. The quantities $M, N_i, C_{ij}$ are functions of the coordinates $x^\alpha=(u, x^i)$, also denoted $\x=(u, \vec x)$, and transform as two-dimensional tensor fields under two-dimensional coordinate transformations $\vec x \mapsto \vec x'(\vec x)$. The \textit{shear} tensor $C_{ij}$ is symmetric and traceless, $q^{ij}C_{ij}=0$. The metric \eqref{Bondi gauge}-\eqref{Bondi expansion} can be shown to solve Einstein's equations $G_{\mu\nu}=8\pi G\, T_{\mu\nu}$ order by order in $1/r$, provided the following constraints are satisfied:
	\begin{align}
	\label{Bondi mass loss}
	\partial_u M&=-\frac{1}{8} N_{ij} N^{ij}+\frac{1}{4} D_i D_j N^{ij}+\frac{1}{8} D_i D^i R_q-4\pi G \lim_{r \to \infty} (r^2\, T_{uu})\,,\\
	\label{angular momentum loss}
	\partial_u N_i&=\partial_i M+\frac{1}{16} \partial_i (C^{mn}N_{mn})-\frac{1}{4} N^{mn}D_i C_{mn}-\frac{1}{4} D_m(C^{mn} N_{in}-N^{mn} C_{in})\\
	\nonumber
	&-\frac{1}{4} D_m D^m D^j C_{ij}+\frac{1}{4} D_m D_i D_n C^{mn}+\frac{1}{4} C_{ij} D^j R_q-8\pi G \lim_{r \to \infty}(r^2 T_{ui}) \,.
	\end{align}
	
	The vacuum metric \eqref{Minkowski in Bondi} is recovered by setting $R_q=2$ (unit round sphere) and $M=N_i=C_{ij}=0$. Even when these functions are nonzero, their contributions in \eqref{Bondi expansion} is subleading in the limit $r\gg 1$. In this sense the corresponding spacetimes are approximately flat away from matter sources. In fact, equations \eqref{Bondi gauge}-\eqref{angular momentum loss} follow from this requirement, together with Einstein's equations and a choice of gauge known as \textit{Bondi gauge}. Detailed derivations may be found in \cite{Tamburino:1966zz,Barnich:2010eb,Compere:2018ylh,Capone:2021ouo}.    
	
	Note also that the asymptotic expansion presented in \eqref{Bondi expansion} assumes \textit{peeling}, i.e., polynomial expansion in $r^{-1}$. It is however known that this property is generically not satisfied, and $\log r$ terms also need to be considered together with an independent $D_{ij}$ tensor at order $O(r^0)$ in the expansion of $g_{ij}$. See \cite{Geiller:2024ryw} and references therein. From the perspective of scattering theory, this is an effect which arises at one loop in relation to long-range gravitational forces between scattered particles \cite{Sahoo:2018lxl}. These aspects still need to be properly incorporated into the framework of Carrollian Holography, and will not be covered in these notes.  
	
\begin{figure}[h!]
		\centering
		\includegraphics[clip,scale=0.5]{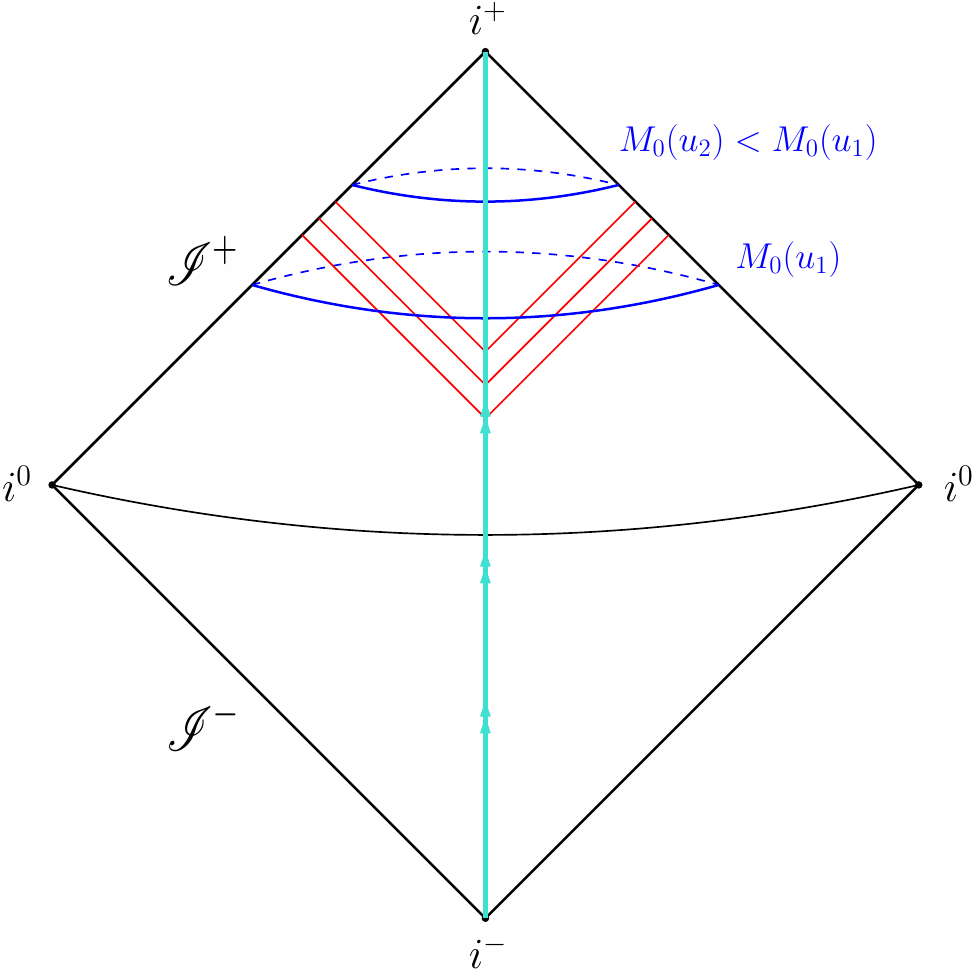}
		\caption{Illustration of the Bondi mass loss. The total mass of the system contained within the Penrose diagram and measured by an observer at $\scri^+$ decreases in retarded time, $M_0(u_2) < M_0(u_1)$. This is because energy is radiated away in form of gravitational waves and other massless fields (red lines) .}
		\label{Penrose2}
	\end{figure}

	The quantities $M,N_i,C_{ij}$ are the first deviations from flat Minkowski space measured by a local asymptotic observer $O(\x)$ hovering at large $r$ and finite $\x=(u,\vec x)$. They capture simple physics: the \textit{Bondi mass aspect} $M(\x)$ roughly corresponds to the Newtonian potential felt by the observer $O(\x)$, i.e., it relates to the mass distribution of the system; the \textit{angular momentum aspect} encodes information about angular momentum and center-of-mass of the system; the \textit{News tensor} $N_{ij}\equiv \partial_u C_{ij}$ carries information about gravitational waves reaching the observer $O(\x)$, its two independent components corresponding to the two possible graviton polarizations. We will define more carefully these concepts in the following sections, momentarily anticipating their physical interpretation. An illuminating example arises when thinking about the total mass of the system measured at retarded time $u$, obtained by integrating the `Newtonian potential' over the celestial sphere $\mathbb{S}^2$,\footnote{This is analogous to measuring total electric charge using Gauss law.}
	\begin{equation}
	M_0(u)\equiv \oint d^2\vec x\, \sqrt{q}\, M(\x)\,. 
	\end{equation}
	Using \eqref{Bondi mass loss}, this total mass is shown to satisfy the evolution equation
	\begin{equation}
	\partial_u M_0(u)=-\frac{1}{8} \oint d^2\vec x\, \sqrt{q}\, N^{ij}N_{ij}-4\pi G \oint d^2\vec x\, \sqrt{q} \lim_{r \to \infty} (r^2 T_{uu})\leq 0\,.
	\end{equation}
	Because matter energy density is positive classically, $T_{uu} \geq 0$, the right-hand side of this equation is negative. The physical interpretation is that mass is radiated away in the form of gravity waves $(N_{ij})$ and other massless fields $(T_{uu})$, and therefore must decrease in retarded time. See Figure~\ref{Penrose2}.
	
	While the physical picture given above is appealing, a critical reader may wonder how it can be justified. They may also be worried about coordinate dependence of the above statements. In the next sections, we will give a purely geometrical description, and justify the interpretation given to the above quantities. We also refer the reader to excellent reviews \cite{Madler:2016xju,Compere:2018aar} where more detailed discussions of the Bondi--Sachs formalism are found.
	
	\subsection{The geometry of null infinity}
	\label{sec:geometry}
	Let us introduce the definition of asymptotic flatness in relation to null infinity $\scri$ and the corresponding induced geometrical data. This approach largely relies on Penrose's conformal compactification \cite{Penrose:1962ij,Penrose:1965am}, which appears extremely well-suited to the description of massless fields and radiation infinitely far away from matter sources. We will mostly follow the treatment given by Geroch  \cite{Geroch1977}, but another useful reference is the review paper of Ashtekar \cite{Ashtekar:2014zsa}.\\ 
	\newline
	\textbf{Definition.} The physical spacetime $(\tmM,\tg_{\mu \nu})$ is said to be \textit{asymptotically locally flat at null infinity} if there exists another spacetime $(\mM,g_{\mu\nu})$ with boundary $\scri$ together with a smooth function $\Omega$ on $\mM$, such that
	\begin{enumerate}
		\item $\tmM$ is diffeomorphic to $\mM-\scri$ (by which they are identified)\,,
		\item  on $\mM - \scri$ : $g_{\mu\nu}=\Omega^2\, \tg_{\mu\nu}$\,,
		\item at $\scri$ : $\Omega=0$,  $\nabla_\mu \Omega\, \hat{\neq}\, 0$ and $\nabla^\mu \Omega\, \nabla_\mu \Omega \hateq 0$\,.
	\end{enumerate}
	The unphysical spacetime $(\mM, g_{\mu\nu})$ is called an \textit{asymptote} of $(\tmM, \tg_{\mu\nu})$. The first condition encodes the idea that $\mM$ is a conformal compactification of $\tmM$. The second condition together with the first part of the third condition states that the conformal boundary $\scri$ is infinitely far away with respect to the physical metric $\tg_{\mu\nu}$. The condition $\nabla_\mu \Omega\, \hat{\neq}\, 0$ ensures that $\Omega$ can be used as `radial' coordinate in a neighborhood of $\scri$, and identifies $n^\mu \equiv g^{\mu\nu} \nabla_\nu \Omega$ with the vector normal to $\scri$. Finally $n^\mu n_\mu \hateq 0$ states that $\scri$ is a null surface with respect to the unphysical metric, which grants it the name \textit{null infinity}. This last condition can actually be derived from Einstein's equations and some minimal assumptions on the falloff rate of the matter stress-energy tensor $\tilde T_{\mu\nu}$ in a neighborhood of~$\scri$. To show this, we first write the relation between the physical and unphysical Einstein tensors,
	\begin{equation}
	\label{Gmunu conformal trans}
	\tilde G_{\mu\nu}=G_{\mu\nu}+2 \Omega^{-1} \nabla_\mu \nabla_\nu \Omega+g_{\mu\nu}\left(3 \Omega^{-2} \nabla^\rho \Omega\, \nabla_\rho \Omega-2\Omega^{-1} \nabla^\rho \nabla_\rho \Omega \right)\,.
	\end{equation}
	Then we make the minimal assumption that $\tilde T_{\mu\nu}$ admits a smooth limit to $\scri$, a condition easily satisfied by massless scalar fields and Maxwell fields for example \cite{Geroch1977}. Multiplying both sides of \eqref{Gmunu conformal trans} by $\Omega$, using Einstein's equations $\tilde G_{\mu\nu}=8\pi G\, \tilde T_{\mu\nu}$ and taking the limit $\Omega \to 0$, we conclude that the quantity 
	\begin{equation}
	f \equiv \Omega^{-1} \nabla^\mu \Omega\, \nabla_\mu \Omega\,,
	\end{equation}
	must also admit a smooth limit to $\scri$. In particular $n^\mu n_\mu \hateq 0$ such that the conformal boundary $\scri$ is a null surface.
	
	The above definition is only concerned with local properties of null infinity. In particular any asymptote for which a portion of $\scri$ has been removed still satisfies this definition. It is then customary to require the global topology of $\scri$ to be $\mathbb{S}^2 \times \mathbb{R}$, in which case $(\tmM,\tg_{\mu\nu})$ is said to be \textit{asymptotically flat at null infinity} \cite{AshtekarSchmidt,Ashtekar:2014zsa}.
	
	The definition of asymptotic flatness leaves significant ambiguity in the choice of conformal factor $\Omega$.
	If $(\mM,g_{\mu\nu},\Omega)$ satisfies the above criteria, so does $(\mM,\Omega',g'_{\mu\nu})$ with
	\begin{equation}
	\label{conformal transformation}
	\Omega' = \omega\, \Omega\,, \qquad g'_{\mu\nu}=\omega^2 g_{\mu\nu}\,,
	\end{equation}
	for $\omega$ any smooth and strictly positive function on $\mM$. Any sensible physical quantity should be independent of this choice, i.e., the Weyl rescaling \eqref{conformal transformation} should be considered a \textit{gauge redundancy}. Said differently, \eqref{conformal transformation} provides an equivalence relation for the asymptotes. There is a unique equivalence class of asymptotes associated with a given asymptotically flat spacetime at null infinity (see theorem 2 in \cite{Geroch1977}). One finds that Weyl rescalings act like
	\begin{equation}
	\begin{split}
	n'^\mu&=\omega^{-1} \left(n^\mu + \Omega\, \nabla^\mu \ln \omega \right)\,,\\
	f'&=\omega^{-1} \left(f+2 n^\mu \nabla_\mu \ln \omega+\Omega\, \nabla_\mu (\ln \omega)\, \nabla^\mu (\ln \omega) \right)\,.
	\end{split}
	\end{equation}
	Using this gauge freedom, it is always possible to choose a conformal frame satisfying the \textit{Bondi condition} 
	\begin{equation}
	\label{Bondi frame}
	f \hateq 0\,,
	\end{equation} 
	in which case Einstein's equations \eqref{Gmunu conformal trans} imply
	\begin{equation}
	\nabla_\mu \nabla_\nu \Omega \hateq 0\,.
	\end{equation}
	As a direct consequence, we also have
	\begin{equation}
	\label{nabla n}
	\nabla_\mu\, n^\nu \hateq 0\,, \qquad \lied_n\, g_{\mu\nu} \hateq 0\,.
	\end{equation}
	The condition \eqref{Bondi frame} leaves a residual rescaling gauge freedom parametrized by functions $\omega >0$ satisfying $\lied_n \omega = n^\mu \nabla_\mu \omega \hateq 0$.
	
	We now have a closer look at the geometrical structure induced on $\scri$. Since the latter is a null surface, it is endowed with the \textit{carrollian} structure $(q_{\alpha\beta}, n^\alpha)$, where $q_{\alpha \beta}$ and $n^\alpha$ are obtained by pullback to $\scri$ of the unphysical metric $g_{\mu\nu}$ and the normal vector $n^\mu$, respectively. The induced metric $q_{\alpha\beta}$ has signature $(0,+,+)$ and $n^\alpha$ spans its kernel, 
	\begin{equation}
	n^\alpha q_{\alpha \beta}=0\,.
	\end{equation}
	Note that under Weyl rescalings, these quantities transform like
	\begin{equation}
	\label{gauge rescaling}
	q'_{\alpha \beta}=\omega^2\, q_{\alpha \beta}\,, \qquad n'^\alpha = \omega^{-1}\, n^\alpha\,.
	\end{equation}
	When the Bondi condition \eqref{Bondi frame} is satisfied, the Levi-Civita derivative operator $\nabla_\mu$ induces a torsionfree and metric compatible derivative operator $D_\alpha$ at $\scri$ satisfying
	\begin{equation}
	\label{Dmu}
	D_\gamma\, q_{\alpha \beta}=0\,, \qquad D_\gamma\, n^\alpha=0\,.
	\end{equation}
	The proof of these statements, together with a detailed discussion of the underlying carrollian affine connection, are given in appendix A of \cite{Nguyen:2022zgs}. In contrast to the case where the metric is non-degenerate, these conditions do not uniquely fix the induced connection. In the present context the connection coefficients left undetermined by \eqref{Dmu} actually encode non-universal information about gravitational radiation \cite{Geroch1977,Ashtekar:1981hw}.  For concreteness, it is sometimes useful to introduce an adapted coordinate system $x^\alpha=(u,x^i)$ in which the metric $q_{\alpha \beta}$ takes the form
	\begin{equation}
	\label{u,x coordinates}
	ds^2=q_{\alpha\beta} \dd x^\alpha \dd x^\beta=0 \dd u^2 +q_{ij} \dd x^i \dd x^j\,, \qquad \partial_u q_{ij}=0\,.
	\end{equation}
	Here $x^i$ are the coordinates covering a `cut' of $\scri$ with topology of the sphere $\mathbb{S}^2$, and $q_{ij}$ is the induced two-dimensional euclidean metric. In this coordinate system $\Gamma^k_{ij}$ and $\Gamma^u_{ij}$ are the only nonzero Christoffel symbols, where the first are the Levi-Civita coefficients associated with the metric $q_{ij}$ while the second are left undetermined by \eqref{Dmu}. 
	
	\paragraph{Relation to Bondi--Sachs metric.} Let us apply this formalism to the metric \eqref{Bondi gauge}, with unit two-dimensional round sphere metric $q_{ij}$ ($R_q=2$). We make the choice of conformal factor $\Omega=1/r$, such that the unphysical metric is
	\begin{align}
	\label{unphysical metric}
	\dd s^2=&- \Omega^2 \dd u^2+2 \dd u \dd \Omega+\left(q_{ij}+\Omega\, C_{ij} \right) \dd x^i \dd x^j+...\,.
	\end{align}
	The normal vector is simply given by
	\begin{equation}
	n_\mu=\delta^\Omega_\mu\,, \qquad n^\mu=\delta^\mu_u+O(\Omega^2)\,,
	\end{equation}
	and the Bondi condition \eqref{Bondi frame} is therefore satisfied. Null infinity $\scri$ is the surface at $\Omega=0$ and is covered with adapted coordinates $x^\alpha=(u,x^i)$. 
	\begin{mytheo}{}{}
		Show that the only nonzero induced connection coefficients at~$\scri$ are \cite{Nguyen:2022zgs}
		\begin{equation}
		\label{Gamma_u}
		\Gamma^k_{ij}=\frac{1}{2}q^{kl}\left(\partial_i q_{lj}+\partial_j q_{il}-\partial_l q_{ij} \right)\,, \qquad \Gamma^u_{ij}=-\frac{1}{2} C_{ij}\,.
		\end{equation}
	\end{mytheo}
	\noindent As promised, the spatial components of the connection are the Levi-Civita coefficients associated with $q_{ij}$. We see that the bulk geometry fixes the otherwise undetermined coefficient $\Gamma^u_{ij}$ to coincide with the shear tensor. The latter encodes dynamical information and hence does \textit{not} constitute universal non-dynamical/background geometrical data. 
	
	\subsection{News and Geroch tensors}
	\label{sec:News}
	Having described the geometry of null infinity, we turn to the quantities needed to properly characterize gravitational radiation. As is well-known, most of the physical information carried by gravitational waves is encoded in the \textit{News tensor} of Bondi and Sachs~\cite{Bondi:1962px,Sachs:1962wk}. In the geometrical formalism used here, its construction was given by Geroch \cite{Geroch1977}.
	
	It turns out that gravitational waves reaching $\scri$ can essentially be described in terms of the unphysical Ricci tensor, or more precisely in terms of the unphysical Schouten tensor
	\begin{align}
	\label{Schouten}
	S_{\mu \nu}&\equiv R_{\mu\nu}-\frac{1}{6} R\, g_{\mu \nu}\,.
	\end{align}
	Indeed its pullback $S_{\alpha\beta}$ to $\scri$ acts as a potential for the leading order Weyl tensor \cite{Geroch1977}. It also satisfies the properties
	\begin{equation}
	S_{\alpha\beta}\, n^\beta=0\,, \qquad S_{\alpha\beta}\, q^{\alpha\beta}=\mathcal{R}\,,
	\end{equation}
	where $\mathcal{R}$ is the scalar curvature of $D_\alpha$, and $q^{\alpha\beta}$ is any `inverse' tensor satisfying $q_{\alpha \gamma} q^{\gamma\delta} q_{\delta \beta}=q_{\alpha\beta}$. 
	The Schouten tensor \eqref{Schouten} is however not gauge invariant, since it transforms under \eqref{conformal transformation} like
	\begin{align}
	S'_{\alpha\beta}&=S_{\alpha \beta}-2 \omega^{-1} D_\alpha D_\beta \omega+4 \omega^{-2} D_\alpha  \omega\, D_\beta \omega- q_{\alpha \beta}\, \omega^{-2} D_\gamma \omega\, D^\gamma \omega\,.
	\end{align}
	Fortunately one can construct a gauge-invariant tensor at $\scri$ in the following elegant way. Geroch taught us that there exists a unique `kinematical' tensor at $\scri$, i.e., constructed out of universal geometrical data, that satisfies \cite{Geroch1977}
	\begin{align}
	\label{Geroch conditions}
	\rho_{[\alpha\beta]}=0\,, \qquad \rho_{\alpha\beta}\, n^\beta=0\,, \qquad \rho_{\alpha \beta}\, q^{\alpha\beta}=\mathcal{R}\,, \qquad D_{[\rho} \rho_{\alpha]\beta}=0\,.
	\end{align}
	The interest of the Geroch tensor lies in its transformation under Weyl rescalings identical to that of $S_{\alpha\beta}$,
	which allows to define the gauge-invariant News tensor
	\begin{equation}
	N_{\alpha\beta}\equiv \rho_{\alpha\beta}-S_{\alpha\beta}\,, \qquad N_{\alpha\beta}\, n^\beta=N_{\alpha\beta}\, q^{\alpha\beta}=0\,.
	\end{equation}
	This is the tensor which characterizes physical gravitational radiation reaching $\scri$. 
	
	\subsubsection*{Relation to Bondi--Sachs metric.} 
	\begin{mytheo}{}{}
		Compute the Schouten tensor \eqref{Schouten} from the metric \eqref{unphysical metric}, and show that
		\begin{equation}
		S_{ij}=q_{ij}-\partial_u C_{ij}\,, \qquad S_{uu}=S_{ui}=0\,, \qquad (\Omega=0)\,.
		\end{equation}
	\end{mytheo}
	\noindent In that case the Geroch tensor is simply $\rho_{ij}=q_{ij}$ and the physical News tensor is thus given by the familiar formula
	\begin{equation}
	N_{ij}=\rho_{ij}-S_{ij}=\partial_u C_{ij}\,.
	\end{equation}
	However, one should be careful not to take this formula outside its regime of validity. In general the Geroch tensor is explicitly needed. See \cite{Nguyen:2022zgs} for more details.

	\subsection{BMS symmetries and charges}
	\label{subsec:BMS}
	We have argued that $\scri$ provides the natural kinematic space to describe physical observables in asymptotically flat spacetimes. As usual we should ask what are the symmetries of this non-dynamical background structure, and what are the corresponding (conserved) charges in the gravitational theory we are currently studying. This will help us give physical meaning to the quantities introduced in Section \ref{subsection: Bondi Sachs}.  
	
	\paragraph{Global BMS symmetries.} The original discovery that asymptotically flat spacetimes possess asymptotic symmetries was made by Bondi, van der Burg, Metzner and Sachs \cite{Bondi:1962px,Sachs:1962wk,Sachs:1962zza}. The corresponding symmetry group is called the \textit{global BMS group}.
	BMS symmetries can be viewed as the subgroup of diffeomorphisms of $\scri$ which preserve the pair $(q_{\alpha\beta},n^\alpha)$ up to a Weyl rescaling \eqref{gauge rescaling}, i.e., up to a gauge transformation \cite{Geroch1977}. Infinitesimally, they are therefore generated by vector fields $\xi^\alpha$ satisfying
	\begin{equation}
	\label{infinitesimal BMS}
	\lied_\xi q_{\alpha\beta}=2 \kappa\, q_{\alpha\beta}\,, \qquad \lied_\xi n^\alpha=-\kappa\, n^\alpha\,.
	\end{equation}
	They must also preserve the Bondi condition $\lied_n q_{\alpha\beta}=0$ which requires the function $\kappa$ to satisfy $\lied_n \kappa=0$. The resulting algebra is also referred to as the level-2 conformal Carroll algebra,  $\mathfrak{bms}=\mathfrak{ccarr}_2$ \cite{Duval:2014uva,Duval:2014lpa}. It has the structure of a semi-direct sum, 
	\begin{equation}
	\mathfrak{bms}=\mathfrak{so}(1,3) \loplus \mathfrak{s}\,.
	\end{equation}
	This is exactly like the Poincaré algebra, except that the translation algebra $\mathfrak{t}=\mathbb{R}^{1,3}$ is replaced by the infinite-dimensional abelian algebra $\mathfrak{s}$. This is not a coincidence as it is understood that Poincaré transformations of Minkowski space induce an action at $\scri$ which is precisely of the type described above \cite{Geroch1977,Oblak:2015qia}. 
	
	To make the symmetry algebra fully explicit, it is best to use adapted coordinates $x^\alpha=(u,x^i)$. A generic vector field solution to \eqref{infinitesimal BMS} then takes the form
	\begin{equation}
	\xi^\alpha=\left(f+\frac{u}{2} D_j Y^j\, ,\, Y^i\right)\,, \qquad \partial_u f=\partial_u Y^i=0\,,
	\end{equation}
	where $Y^i$ is a two-dimensional conformal Killing vector field satisfying
	\begin{equation}
	\label{Killing equation}
	D_i Y_j+D_j Y_i=D_k Y^k\, q_{ij}\,.
	\end{equation}
	\begin{mytheo}{}{}
		Show that these vector fields close under the Lie bracket
		\begin{align}
		\label{bms algebra}
		\hat{\xi}_{12}^\alpha= [\xi_1,\xi_2]^\alpha\,,
		\end{align}
		with
		\begin{equation}
		\label{bms algebra bis}
		\begin{split}
		f_{12}&=Y_1^i D_i f_2+\frac{1}{2}f_1 D_i Y_2^i\ - (1 \leftrightarrow 2)\,,\\
		Y^i_{12}&=Y_1^j D_j Y_2^i-(1 \leftrightarrow 2)\,.
		\end{split}
		\end{equation}
	\end{mytheo}
	\noindent As is well-known, globally well-defined solutions to \eqref{Killing equation} generate the Lie algebra $\mathfrak{so}(1,3)$. The vector fields $\xi=f(\vec x)\partial_u$ generate the \textit{supertranslation} algebra $\mathfrak{s}$. The translation subalgebra $\mathfrak{t}\in \mathfrak{s}$ corresponds to the four lowest spherical harmonics of $f(\vec x)$ \cite{Geroch1977,Ashtekar:2014zsa}. For a more detailed account, the interested reader may also consult \cite{Nguyen:2022zgs}.
	
	\paragraph{Global BMS charges.} Background symmetries usually imply the existence of conserved charges, which would also act as symmetry generators in a Hamiltonian formulation of the theory. The charges associated with BMS symmetries are of the form \footnote{Many inequivalent expressions can be found in the literature, which differ by terms vanishing in the limit $u \to -\infty$.}
	\begin{equation}
	\label{BMS charges}
	\begin{split}
	Q[f](u)&=\oint_{\mathbb{S}_u} \dd^2 \vec x\, \sqrt{q}\, f(\vec x)\, M(u,\vec x)\,,\\
	Q[Y](u)&=\oint_{\mathbb{S}_u} \dd^2 \vec x\, \sqrt{q}\, Y^i(\vec x)\, N_i(u,\vec x)\,.\\
	\end{split}
	\end{equation}
	These charges are defined on a cut of $\scri^+$, i.e., the celestial sphere at constant value of the retarded time $u$. At first sight, this appears analogous to conserved charges defined over constant time-slices $\Sigma_t$ in conventional field theories. Importantly though, the charges \eqref{BMS charges} are \textit{not} conserved in retarded time, as easily shown using \eqref{Bondi mass loss}-\eqref{angular momentum loss}. This is because, from a bulk spacetime perspective, $\mathbb{S}_u$ is not the boundary of a Cauchy slice $\Sigma_t$. Data is lost in the form of gravitational waves ($N_{\alpha \beta})$ and massless matter ($T_{\alpha\beta}$) as one evolves along $u$. See Figure~\ref{Penrose2}. This can be fixed by taking the limit $u \to -\infty$ where $\scri^+$ `joins' spatial infinity $i^0$. In that limit, radiation vanishes such that \eqref{BMS charges} become genuine conserved charges. It is also in a neighborhood of $i^0$ that they can be promoted to canonical symmetry generators. See \cite{Capone:2022gme} and references therein. 
	
	Because $Q[Y](u)$ are charges associated with Lorentz symmetries $\operatorname{SO}(1,3) \subset \operatorname{ISO}(1,3) \subset \operatorname{BMS}$, we identify them with angular momentum and center-of-mass of the gravitational system. Similarly, total energy corresponds to $M_0(u)=Q[f_0](u)$ with constant parameter $f_0(\vec x)=f_0$, while three-momentum $P_m(u)=Q[f_m](u)$ is obtained from the three $\ell=1$ spherical harmonics $f_m(\vec x)=Y_m^{\ell=1}(\vec x)$. Higher spherical harmonics yield supertranslation charges which do not have non-gravitational counterparts. Although this identification can only be made precise in a neighborhood of spatial infinity $i^0$, the physical interpretation of \eqref{BMS charges} is often retained for arbitrary values of retarded time $u$.

	\paragraph{Extended BMS symmetries.} An extension of the BMS algebra was proposed by Barnich and Troessaert \cite{Barnich:2009se,Barnich:2010eb}. Their proposal is simply to consider \textit{local} solutions of the conformal Killing equation \eqref{Killing equation}. In complex stereographic coordinates $z=x^1+ix^2,\zbar=x^1-ix^2$ covering the celestial sphere $\mathbb{S}^2$, the corresponding vector fields have (anti)-meromorphic components $\xi^z(z)$ and $\xi^{\zbar}(\zbar)$. Therefore the extended BMS algebra takes the form
	\begin{equation}
	\mathfrak{ebms}=\left[ \mathfrak{diff}(\mathbb{S}^1) \oplus \mathfrak{diff}(\mathbb{S}^1)  \right] \loplus \mathfrak{s}^*\,.
	\end{equation}
	Note that $\mathfrak{diff}(\mathbb{S}^1)$ is also the centerless Virasoro algebra and it is often this latter terminology that is used. The \textit{superrotation} vector fields sitting in the quotient $\left[ \mathfrak{diff}(\mathbb{S}^1) \oplus \mathfrak{diff}(\mathbb{S}^1)  \right]/\mathfrak{so}(3,1)$ are not globally well-defined since they have poles at isolated points of the sphere~$\mathbb{S}^2$. It might therefore look like these additional transformations should not be allowed if we restrict to everywhere smooth metric fields $q_{ij}$ and cuts of $\scri$ with sphere topology. The situation is identical to that of two-dimensional CFTs, and the existence of these non-global symmetries is sufficient to guarantee the existence of locally conserved currents \cite{Belavin:1984vu}. This justifies the extended version of the BMS algebra. Finally the supertranslation algebra $\mathfrak{s}^*$ is generated by functions $f$ that can also have singularities on the sphere, in order for the Lie algebra \eqref{bms algebra bis} to close, and it is therefore larger than the algebra $\mathfrak{s}$ of smooth supertranslations. 
	
	\paragraph{Generalized BMS symmetries.} A further generalization of the asymptotic symmetry algebra has been proposed by Campiglia and Laddha \cite{Campiglia:2014yka,Campiglia:2015yka}. Generalized BMS symmetries preserve the conformal class $(\epsilon_{\alpha\beta\gamma},n^\alpha)$ where $\epsilon_{\alpha\beta\gamma}$ is the volume form at $\scri$, together with the Bondi condition $\lied_n q_{\alpha\beta}=0$. It will not play a role in these lectures.
	
	\section{Scattering states and carrollian conformal fields}
	\label{sec: states and fields}
	
	We have argued in the previous section that null infinity $\scri$ is the natural kinematic space over which to define classical observables. These include the mass aspect $M(\x)$, the angular momentum aspect $N_i(\x)$, the News $N_{ij}(\x)$ as well as other forms of massless radiation fields. BMS symmetries, viewed as conformal isometries of $\scri$, allow to identify notions of (super)-momentum, angular momentum, and center-of-mass of the total gravitating system which are directly built from these quantities.   
	
	We will now move from the classical theory to the quantum theory. More specifically, we will consider a quantum theory of gravitational scattering where the bulk spacetime is allowed to fluctuate strongly. As a concrete physical setup, one can imagine a high-energy collision leading to black hole formation at intermediate times, and to evaporation into Hawking radiation in the asymptotic future. This situation cannot be described by perturbative QFT in Minkowski spacetime, which is the usual basis of scattering theory. But even if spacetime dynamically develops large local curvature, under the assumption of asymptotic flatness, null infinity $\scri$ always provides a non-dynamical background structure on which to anchor physical observables. 
	
	Following conventional wisdom, we should try to define a quantum theory whose Hilbert space admits a \textit{unitary} action of the symmetry group \cite{Weinberg:1995mt}. It would therefore appear natural to discuss unitary irreducible representations (UIRs) of the BMS group as building blocks for the Hilbert space of our quantum gravitational theory. Although work in this direction has been carried out \cite{Sachs-1962,Mccarthy:1972ry,McCarthy1975TheBG,Longhi:1997zt,Bekaert:2024uuy,Bekaert:2025kjb,Donnay:2026urd,Ruzziconi:2026isv}, its connection to relativistic scattering theory is not yet well developed. For the time being, we will restrict our attention to its Poincar\'e subgroup $\operatorname{ISO}(1,3) \subset \operatorname{BMS}$, and consider a Hilbert space made of Poincar\'e UIRs. In this way, we adopt the same foundations as relativistic scattering theory \cite{Weinberg:1995mt}, which will make it easier for everyone to follow the subsequent developments. 
	
	The punchline of this section is the following: massless particles states, including photons and gravitons, can be encoded in conformal fields defined \textit{locally} at $\scri$. First, we will review Wigner's construction of massless particle states using a judicious parametrization of the group elements \cite{Nguyen:2023vfz}. Second, we will independently construct conformal field representations of $\operatorname{ISO}(1,3)$ \cite{Bagchi:2016bcd,Nguyen:2023vfz} in a way that parallels the standard construction of $\operatorname{SO}(2,4)$ conformal fields \cite{Mack:1969rr}. We will then provide the intertwining map between those two seemingly distinct classes of representations. We will also take this opportunity to introduce the \textit{celestial basis} of states, which involves decomposing Poincar\'e UIRs into Lorentz UIRs \cite{Banerjee:2018gce,Iacobacci:2024laa}. They provide the basis of Celestial Holography \cite{Pasterski:2021raf,McLoughlin:2022ljp,Donnay:2023mrd}.
	
	We will end this section with a preliminary discussion of the equivalence between massless scattering amplitudes and correlation functions of the carrollian conformal fields, which we will return to in Section~\ref{sec: carrollian correlators and amplitudes}. 
	
	\subsection{Poincar\'e algebra}
	The Poincar\'e algebra $\mathfrak{iso}(1,3)$ is usually given in the form
	\begin{equation}
	\begin{split}
	\left[\tilde J_{\mu\nu},\tilde J_{\rho \sigma}\right]&=-i\left(\eta_{\mu\rho} \tilde J_{\nu \sigma}+\eta_{\nu \sigma}\tilde J_{\mu \rho}-\eta_{\mu\sigma} \tilde J_{\nu\rho}-\eta_{\nu \rho} \tilde J_{\mu \sigma} \right)\,,\\
	\left[\tilde J_{\mu\nu}, \tilde P_\rho \right]&=-i\left(\eta_{\mu\rho}\tilde P_\nu-\eta_{\nu \rho}\tilde P_\mu \right)\,.
	\end{split}
	\end{equation}
	While this basis arises most naturally in relation to isometries of Minkowski spacetime, for our purposes it is useful to split the indices $\mu=\{0,i,3\}$ with $i=\{1,2\}$ and introduce the alternative basis of generators
	\begin{equation}
	\label{tilde J}
	\tilde J_{ij}=J_{ij}\,, \qquad \tilde J_{0i}=\frac{1}{2}\left(P_i+K_i \right)\,, \qquad \tilde J_{i3}= \frac{1}{2}\left(P_i-K_i \right)\,, \qquad \tilde J_{03}=-D\,,
	\end{equation}
	and
	\begin{equation}
	\label{tilde P}
	\tilde P_0=\frac{1}{\sqrt{2}}(H+K)\,, \qquad \tilde P_i=-\sqrt{2}\, B_i\,, \qquad \tilde P_3=\frac{1}{\sqrt{2}}(K-H)\,.
	\end{equation}
	In this new basis the algebra relations are given by
	\begin{align}
	\nonumber
	\left[J_{ij}\,,J_{mn}\right]&=-i\left(\delta_{im} J_{jn}+\delta_{jn} J_{im}-\delta_{in} J_{jm}-\delta_{jm} J_{in} \right)\,, & \left[D\,, P_i \right]&=i P_i\,,\\
	\nonumber
	\left[J_{ij}\,,P_k \right]&=-i \left(\delta_{ik} P_j-\delta_{jk} P_i \right)\,, & \left[D\,, H \right]&=i H\,,\\
	\nonumber
	\left[J_{ij}\,,K_k \right]&=-i \left(\delta_{ik} K_j-\delta_{jk} K_i \right)\,, & \left[D\,, K_i \right]&=-i K_i\,,\\
	\label{conformal Carroll algebra}
	\left[J_{ij}\,,B_k \right]&=-i\left(\delta_{ik}B_j-\delta_{jk} B_i\right)\,, & \left[D\,, K \right]&=-i K\,,\\
	\nonumber
	\left[B_i\,, P_j \right]&=i \delta_{ij} H\,, & \left[H\,, K_i\right]&=2i B_i\,,\\
	\nonumber
	\left[B_i\,,K_j \right]&=i \delta_{ij} K\,, & \left[K\,, P_i\right]&=2i B_i\,,\\
	\nonumber
	\left[K_i\,, P_j\right]&=-2i\left(\delta_{ij} D-J_{ij} \right)\,,
	\end{align}
	with the remaining commutators being zero.
	
	This new basis is most natural from the perspective of $\scri$. Indeed, let us consider the flat metric representative
	\begin{equation}
	ds^2_{\scri}=q_{\alpha\beta} \dd x^\alpha \dd x^\beta=0\, \dd u^2 +\delta_{ij} \dd x^i \dd x^j\,.
	\end{equation}
	It is in the same conformal class as the metric \eqref{u,x coordinates} we considered earlier (to see this, choose $x^i$ in \eqref{u,x coordinates} to be stereographic coordinates).
	
	\begin{mytheo}{}{}
		Show that the vector fields solution to \eqref{infinitesimal BMS} which generate $\mathfrak{iso}(1,3) \subset \mathfrak{bms}$, may be written
		\begin{equation}
		\label{xi Poincare}
		\begin{split}
		\xi^u&=a+b_i\, x^i+k\, x^2+(\lambda-2 k_i x^i)u\,,\\
		\xi^i&=a^i+\omega^i{}_j\, x^j+\lambda x^i+k^i x^2-2k_j x^j x^i\,.
		\end{split}
		\end{equation}
	\end{mytheo}
	\noindent This can be neatly rewritten
	\begin{align}
	\label{infinitesimal coordinate change}
	x'^{\alpha}&\equiv x^\alpha+\xi^\alpha=x^\alpha+i(a H+a^i P_i+\frac{1}{2} \omega^{ij} J_{ij}+ b^i B_i+ \lambda D+ k K+ k^i K_i)\, x^{\alpha}\,,
	\end{align}
	provided we introduce the differential operators
	\begin{align}
	\nonumber
	P_i&=-i \partial_i\,, & J_{ij}&=i(x_i \partial_j-x_j \partial_i )\,,\\
	\label{Carroll generators}
	D&=-i(u\partial_u+x^i\partial_i )\,, & K_i&=-i(x^2 \partial_i-2x_i(u\partial_u+x^j\partial_j))\,, & K&=-i x^2 \partial_u\,,\\
	\nonumber
	H&=-i \partial_u\,, & B_i&=-i x_i \partial_u\,.
	\end{align}
	\begin{mytheo}{}{}
		Show that the operators \eqref{Carroll generators} satisfy the algebra relations \eqref{conformal Carroll algebra}.
	\end{mytheo}
	\noindent Thus, we found a new `coordinate realization' of the Poincar\'e symmetries. The kinematic space they act upon is $\scri$ rather than $\mathbb{M}^4$. Crucially, this representation is such that the quadratic Casimir operator vanishes identically,
	\begin{equation}
	\label{C2=0}
	\mathcal{C}_2=\tilde P^\mu \tilde P_\mu=-\tilde P_0^2+\tilde P^i \tilde P_i+\tilde P_3^2=-(HK+KH)+2B^i B_i=2x^2 \partial_u^2-2x^2 \partial_u^2=0\,.
	\end{equation}
	This is in stark contrast with the standard representation $\mathcal{C}_2=\partial^\mu \partial_\mu$ associated with Minkowski space $\mathbb{M}^{4}$. This is a hint that local fields at $\scri$ will describe massless particles only.\\
	
	\textit{Remark.} The coordinate representation \eqref{Carroll generators} can be obtained from an ultra-relativistic limit ($c \to 0$) of the conformal coordinate transformations in $\mathbb{M}^3$. Indeed, in that limit the metric of $\mathbb{M}^3$ become that of $\scri$, and one can implement the algebra contraction $\mathfrak{so}(2,3) \to \mathfrak{iso}(1,3)$. See the appendix of \cite{Nguyen:2023vfz} for more details. If one only considers the isometries of $\mathbb{M}^3$ rather than its conformal isometries, the ultra-relativistic contraction yields the global carroll algebra, $\mathfrak{iso}(1,2) \to \mathfrak{carr}(3)$. This was first studied by Levy--Leblond \cite{Levy1965}, who coined the term `carrollian' (i.e., ultra-relativistic) in reference to an episode of Lewis Carroll's \textit{Through the looking-glass}, where Alice and The Queen run as fast as they can without moving anywhere, illustrating the physical implication of the limit $c \to 0$. 
	
	\subsection{Massless particle states}
	Let us now describe massless particles as UIRs of $\operatorname{ISO}(1,3)$. Following Wigner's method of induced representation \cite{Wigner1931,Weinberg:1995mt}, massless particle states are defined starting from the reference null momentum
	\begin{equation}
	k^\mu=\frac{1}{\sqrt{2}}(1,\vec 0,1)\,,
	\end{equation}
	together with a UIR of the massless little group $\operatorname{ISO}(2) \subset \operatorname{SO}(1,3)$ that leaves this momentum invariant. 
	\begin{mytheo}{}{}
		Show that the massless little group is generated by $\{ J_{ij},K_i \}$ as defined in \eqref{tilde J}.
	\end{mytheo}
	\noindent We can characterize this inducing represensation by
	\begin{align}
	\label{massless little group rep}
	J_{ij}\, |k\rangle_\sigma=(\Sigma^{(s)}_{ij})_\sigma{}^{\sigma'}|k\rangle_{\sigma'}\,, \qquad K_i\, |k \rangle_\sigma=0\,, \qquad \tilde P_\mu\, |k \rangle_\sigma=k_\mu\, |k\rangle_\sigma\,,  
	\end{align}
	where $\Sigma^{(s)}_{ij}$ is a hermitian matrix furnishing a spin-$s$ representation of $\operatorname{SO}(2)$ or $\operatorname{Spin}(2)$. Note that the operator $\tilde P_\mu$ above is the standard momentum generator, not to be confused with the boost generators $P_i$ defined in \eqref{tilde P}. It is often useful to work with states of definite helicity, in which case we replace the first equation by
	\begin{equation}
	J_{ij} |k\rangle_J=J \varepsilon_{ij}|k\rangle_J\,, \qquad J=\pm s\in \frac{1}{2}\mathbb{Z}\,. 
	\end{equation}
	A generic momentum state is then obtained by applying the remaining boost generators,
	\begin{equation}
	\label{generic massless momentum state}
	|p(\omega,\vec x)\rangle_\sigma \equiv e^{ix^i P_i}\, e^{i\ln \omega\, D}\, |k\rangle_\sigma\,.
	\end{equation}
	The last equation is just one specific way to implement Wigner's induced representation, resulting in the specific momentum parametrisation  
	\begin{equation}
	\label{massless momentum parametrisation}
	p^\mu(\omega,\vec x)=\omega\, q^\mu(\vec x)\,, \qquad q^\mu(\vec x)\equiv \frac{1}{\sqrt{2}} (1+x^2,2\vec x,1-x^2)\,.
	\end{equation}
	The action of the Lorentz generators on the generic state \eqref{generic massless momentum state} can be worked out using the algebra relations together with \eqref{massless little group rep}-\eqref{generic massless momentum state}, yielding
	\begin{equation}
	\label{massless rep}
	\begin{split}
	P_i\, |p\rangle &=-i \partial_i\, |p \rangle\,,\\
	J_{ij}\, |p\rangle&=-i\left(x_i \partial_j-x_j\partial_i+i \Sigma_{ij} \right) |p\rangle\,,\\
	D\, |p\rangle&=i\left(-\omega \partial_\omega+x^i\partial_i \right)|p\rangle\,,\\
	K_i\, |p\rangle &=i\left(-2x_i \omega \partial_\omega+2x_i x^j \partial_j-x^2 \partial_i+2i x^j \Sigma_{ij} \right)|p\rangle\,,
	\end{split}
	\end{equation}
	where spin indices have been suppressed for convenience.

	A generic state in the corresponding one-particle Hilbert space takes the form
	\begin{equation}
	\label{generic massless state}
	|\psi \rangle=\sum_\sigma \int [d^3  p(\omega,\vec x)]\, \psi_\sigma(\omega,\vec x) |p(\omega,\vec x)\rangle_\sigma\,,
	\end{equation}
	with $\psi_\sigma(\omega,\vec x)$ any complex wavefunction normalizable with respect to the inner product
	\begin{equation}
	\label{massless inner product}
	\langle \phi | \psi \rangle=\sum_\sigma \int [d^3 p(\omega,\vec x)]\, \phi_\sigma(\omega,\vec x)^*\, \psi_\sigma(\omega,\vec x)\,,    
	\end{equation}
	and with $[d^3 p(\omega,\vec x)]$ the Lorentz-invariant measure on the forward null momentum cone,
	\begin{equation}
	[d^3 p(\omega,\vec x)]= \omega \dd\omega \dd^2\vec x\,.
	\end{equation}
	We note that in the limit of vanishing frequency $\omega$, square integrability of the wavefunctions requires 
	\begin{equation}
	\label{zero frequency behavior}
	\psi_\sigma(\omega,\vec x)=o\left(\omega^{-1}\right)\,, \qquad (\omega \to 0)\,.
	\end{equation}
	Additionally, even though the states $|p(\omega,\vec x)\rangle_\sigma$ do not strictly belong to this Hilbert space, one can assign them the distributional inner product 
	\begin{equation}
	\label{inner product Wigner}
	{}_{\sigma'}\langle p_1|\,p_2\rangle_\sigma=(\omega_1)^{-1}\, \delta(\omega_1-\omega_2) \delta(\vec x_1-\vec x_2) \delta_{\sigma\sigma'}\,,
	\end{equation}
	consistently with \eqref{generic massless state}-\eqref{massless inner product}.
	
	\subsubsection*{The celestial basis}
	Although this is not the main focus of these lectures, let us take this opportunity to introduce the \textit{celestial basis} of the Hilbert space, which lies at the basis of Celestial Holography \cite{Pasterski:2021raf,McLoughlin:2022ljp,Donnay:2023mrd}. The idea here is to decompose particle states into irreps of the Lorentz group. Because the Poincar\'e group acts unitarily, the same is true for its Lorentz subgroup. We will treat massless particles, and we refer to \cite{Iacobacci:2024laa} for a general treatment including spinning massive particles.
	
	In practice, we want to decompose the massless wavefunctions $\psi_\sigma(\omega,\vec x)$ over a set of wavefunctions $\psi_\sigma(\Delta^*,\vec x)$ carrying irreducible representations of $\operatorname{SO}(1,3)$. This can be achieved by noticing that the Lorentz transformations \eqref{massless rep} are the same as those of a two-dimensional conformal field provided we make the replacement $\Delta^* \leftrightarrow -\omega \partial\omega$. This is implemented through the Mellin transform
	\begin{equation}
	\label{Mellin transform}
	\psi_\sigma(\Delta^*,\vec x)=\frac{1}{\sqrt{2\pi}} \int_0^\infty d\omega\, \omega^{\Delta^*-1}\, \psi_\sigma(\omega,\vec x)\,.
	\end{equation}
	The condition \eqref{zero frequency behavior} implies that $\psi_\sigma(\Delta^*,\vec x)$ is holomorphic in $\Delta^*$ on the complex half-plane $\text{Re}(\Delta^*)\geq 1$. The inverse Mellin transform then provides the sought-for decomposition of the massless wavefunction into Lorentz irreps,
	\begin{equation}
	\label{decomposition massless wavefunction}
	\psi_\sigma(\omega,\vec x)=\frac{1}{\sqrt{2\pi} }\int_{-\infty}^{\infty} d\nu\, \omega^{-1+i\nu}\, \psi_\sigma(\Delta^*_\nu,\vec x)\,,  \qquad \Delta^*_\nu=1-i\nu\,.
	\end{equation}
	This can be translated to a decomposition at the level of basis states,
	\begin{equation}
	\label{decomposition p}
	|p(\omega,\vec x)\rangle_\sigma=\frac{1}{\sqrt{2\pi}}\int_{-\infty}^{\infty} d\nu\, \omega^{-1-i\nu}\, |\nu,\vec x\rangle_\sigma\,, 
	\end{equation}
	with inverse
	\begin{equation}
	\label{inverse p decomposition}
	|\nu,\vec x\rangle_\sigma=\frac{1}{\sqrt{2\pi}}\int_0^\infty d\omega\, \omega^{i\nu} |p(\omega,\vec x)\rangle_\sigma\,.     
	\end{equation}
	The states $|\nu, \vec x\rangle_\sigma$ transform as two-dimensional conformal fields with scaling dimension $\Delta=1+i\nu$, a value corresponding to \textit{unitary} representations of $\operatorname{SO}(1,3)$ in the so-called \textit{continuous principal series} \cite{Dobrev:1977qv,Sun:2021thf}. The inner product for these states is given by
	\begin{equation}
	\label{inner product}
	{}_{\sigma'}\langle \nu', \vec y\, |\, \nu, \vec x \rangle_\sigma= \delta(\nu-\nu') \delta(\vec x-\vec y)\, \delta_{\sigma \sigma'} \,.
	\end{equation}
	\begin{mytheo}{}{}
		Show that \eqref{inner product} is consistent with \eqref{inner product Wigner} and \eqref{inverse p decomposition}.
	\end{mytheo}
	
	\subsection{Carrollian conformal fields}
	\label{subsec: carrollian fields}
	We now turn to the construction of fields defined locally on $\scri$ and transforming covariantly under $\operatorname{ISO}(1,3)$\cite{Bagchi:2016bcd,Nguyen:2023vfz}. Like Mack and Salam did for relativistic conformal fields in $\mathbb{M}$ \cite{Mack:1969rr}, we first look for finite-dimensional irreps of the isotropy subgroup $H$ leaving the origin $\x=0$ invariant. Looking at \eqref{Carroll generators} we see that the latter is generated by 
	\begin{equation}
	H=\text{span}(J_{ij}, B_i,K_i,K,D )\,.
	\end{equation}
	As before, we consider that our field $O(0)$ transforms under rotation like
	\begin{equation}
	\label{J B actions}
	\left[J_{ij}, O(0)\right]=J \varepsilon_{ij}\, O(0)\,, \qquad J=\pm s\,.
	\end{equation}
	Since $B_i$ and $K_i$ are $SO(2)$ vectors, finite dimensionality of $O(0)$ requires them to act trivially,
	\begin{equation}
	\label{Bi Ki action}
	\left[B_i,O(0)\right]=\left[K_i,O(0)\right]=0\,.
	\end{equation}
	Consistency with the algebra \eqref{conformal Carroll algebra} then requires the generator $K$ to act trivially too,
	\begin{equation}
	\label{K action}
	\left[B_i,K_j\right]=i\delta_{ij} K \quad \Rightarrow \quad \left[K,O(0)\right]=0\,.
	\end{equation}
	On the other hand, since it commutes with the rotation generator, the action of the dilation operator can be diagonalised,
	\begin{equation}
	\label{action D}
	\left[D, O(0)\right]=i\Delta\, O(0)\,, 
	\end{equation}
	with $\Delta$ an arbitrary complex number at this point. With this we have fully specified the action of the isotropy group on $O_{\Delta,J}(0)$, the local field at the origin.
	\begin{mytheo}{}{}
		Show that $K_\alpha=(K,K_i)$ and $P_\alpha=(H,P_i)$ act as lowering and raising operator for the conformal dimension, 
		\begin{equation}
		\begin{split}
		\left[D,\left[K_\alpha, O(0)\right] \right]&=i(\Delta-1)\left[K_\alpha, O(0) \right]\,,\\
		\left[D,\left[P_\alpha, O(0)\right] \right]&=i(\Delta+1)\left[P_\alpha, O(0) \right]\,.
		\end{split}
		\end{equation}
	\end{mytheo}
	\noindent We then induce the full Poincar\'e representation by `translating' the fields,
	\begin{equation}
	\label{translated fields}
	O_{\Delta,J}(\x)\equiv U(\x)\, O_{\Delta,J}(0)\, U(\x)^{-1}\,, 
	\end{equation}
	with the group elements
	\begin{equation}
	U(\x)=e^{-i x^\alpha P_\alpha}=e^{-i(uH+x^i P_i)}\,.
	\end{equation}
	This definition directly implies
	\begin{equation}
	[P_\alpha\,,O_{\Delta,J}(\x)]=i\partial_\alpha O_{\Delta,J}(\x)\,.
	\end{equation}
	To work out the action of one of the isotropy generators $X$ on the translated field $O_{\Delta,J}(\x)$, we make use of the identity
	\begin{equation}
	[X\,,\psi_i(\x)]=U(\x) [X'\,,\psi_i\,] U(\x)^{-1}\,,    
	\end{equation}
	where
	\begin{equation}
	X'=U(\x)^{-1}XU(\x)=\sum_{n=0}^\infty \frac{i^n}{n!} x^{\alpha_1}...\,x^{\alpha_n}\, [P_{\alpha_1}\,,[...\,,[P_{\alpha_n}\,,X]]]\,.
	\end{equation}
	\begin{mytheo}{}{}
		Using the algebra relations \eqref{conformal Carroll algebra}, show that
		\begin{equation}
		\label{translated generators}
		\begin{split}
		J_{ij}'&=J_{ij}-x_i P_j+x_j P_i\,,\\
		D'&=D+uH+x^iP_i\,,\\
		K'&=K+2x^i B_i+x^2 H\,,\\
		K_i'&=K_i-2uB_i-2x_i D+2x^jJ_{ij}-2x_iuH-2x_ix^jP_j+x^2P_i\,,\\
		B_i'&=B_i+x_iH\,.
		\end{split}
		\end{equation}
	\end{mytheo}
	\noindent Hence we have, for instance,
	\begin{equation}
	\begin{split}
	[D\,,O_{\Delta,J}(\x)]&=U(\x) [D+x^\alpha P_\alpha\,,O_{\Delta,J}] U(\x)^{-1}\\
	&=U(\x) \left(i\Delta\, O_{\Delta,J}+x^\alpha[P_\alpha\,,O_{\Delta,J}]\right) U(\x)^{-1}\\
	&=i\left(\Delta+x^\alpha\partial_\alpha\right)O_{\Delta,J}(\x)\,,\\
	\end{split}
	\end{equation}
	where in the last line we made use of
	\begin{equation}
	\begin{split}
	U(\x)[P_\alpha\,,O_{\Delta,J}]U(\x)^{-1}&=U(\x)P_\alpha\, O_{\Delta,J} U(\x)^{-1}-U(\x) O_{\Delta,J} P_\alpha U(\x)^{-1}\\
	&=P_\alpha\, U(\x) O_{\Delta,J} U(\x)^{-1}-U(\x) O_{\Delta,J} U(\x)^{-1} P_\alpha\\
	&=[P_\alpha\,,O_{\Delta,J}(\x)]=i\partial_\alpha O_{\Delta,J}(\x)\,.
	\end{split}
	\end{equation}
	Similar manipulations can be performed for the remaining generators, yielding \cite{Nguyen:2023vfz}
	\begin{equation}
	\label{Carrollian induced rep}
	\begin{split}
	\left[P_\alpha, O_{\Delta,J}(\x)\right]&=i \partial_\alpha O_{\Delta,J}(\x)\,,\\
	\left[J_{ij}, O_{\Delta,J}(\x)\right]&=i(-x_i \partial_j+x_j\partial_i -iJ \varepsilon_{ij})\,O_{\Delta,J}(\x)\,,\\
	\left[D, O_{\Delta,J}(\x)\right]&=i(\Delta+u\partial_u+x^i\partial_i )\,O_{\Delta,J}(\x)\,,\\
	\left[K, O_{\Delta,J}(\x)\right]&=i x^2 \partial_u O_{\Delta,J}(\x)\,,\\
	\left[K_i, O_{\Delta,J}(\x)\right]&=i(-2x_i \Delta-2iJ x^j \varepsilon_{ij}-2x_i u\partial_u-2x_i x^j \partial_j+x^j x_j \partial_i )\,O_{\Delta,J}(\x)\,,\\
	\left[B_i, O_{\Delta,J}(\x)\right]&=ix_i\partial_u O_{\Delta,J}(\x)\,.
	\end{split}
	\end{equation}
	
	\begin{mytheo}{}{}
		Show that \eqref{Carrollian induced rep} exactly correspond to the infinitesimal transformation of a tensor density,
		\begin{equation}
		\label{carrollian tensor}
		\delta O_{\Delta,J}=\left(\lied_\xi -\Delta\, \Omega_\xi \right) O_{\Delta, J}\,, \qquad \Omega_\xi=\frac{1}{3} \partial_\alpha \xi^\alpha\,,
		\end{equation}
		for $\xi$ is a vector field of the form \eqref{xi Poincare}. Details can be found in \cite{Nguyen:2023vfz}.
	\end{mytheo}
	\noindent A field transforming as \eqref{Carrollian induced rep} is called a \textit{carrollian conformal primary field}. We have the important following property, given as an exercise:
	\begin{mytheo}{}{}
		For $O_{\Delta,J}$ a primary field, show that $\partial_u O_{\Delta,J}$ is also a primary field, with shifted scaling dimension $\Delta'=\Delta+1$.    
	\end{mytheo}
	\noindent This is to be contrasted with conventional CFT, where derivatives of primary fields do not transform as primary fields.
	
	\paragraph{Complex coordinates.}
	It is often customary to use complex stereographic coordinates on the celestial sphere,
	\begin{equation}
	\label{z zbar}
	z=x^1+ix^2\,, \qquad \zbar=x^1-i x^2\,.
	\end{equation}
	In that case it is more natural to define the generators~\cite{Bagchi:2023cen}
	\begin{equation}
	\begin{aligned}
	P_{-1,-1}&=-iH\,, &\quad  L_{-1}&=-\frac{i}{2}(P_1-iP_2)\,, &\quad  \bar L_{-1}&=-\frac{i}{2}(P_1+iP_2)\,,\\
	P_{0,-1}&=-i(B_1+iB_2)\,, &\quad L_0&=-\frac{i}{2}(D+iJ_{12})\,, &\quad \bar L_0&=-\frac{i}{2}(D-iJ_{12})\,,\\
	P_{-1,0}&=-i(B_1-iB_2)\,, &\quad L_1&=\frac{i}{2}(K_1+iK_2)\,, &\quad \bar L_1&=\frac{i}{2}(K_1-iK_2)\,,\\
	P_{0,0}&=-iK
	\end{aligned}
	\end{equation}
	such that \eqref{Carrollian induced rep} become
	\begin{align}
	\label{complex field transformations}
	\begin{split}
	\left[P_{-1,-1}\,,O_{\Delta,J}(\x) \right]&=\partial_u O_{\Delta,J}(\x)\,,\\
	\left[P_{0,-1}\,,O_{\Delta,J}(\x)\right]&=z\partial_u O_{\Delta,J}(\x)\,,\\
	\left[P_{-1,0}\,,O_{\Delta,J}(\x)\right]&=\zbar \partial_u O_{\Delta,J}(\x)\,,\\
	\left[P_{0,0}\,,O_{\Delta,J}(\x)\right]&=z\zbar \partial_u O_{\Delta,J}(\x)\,,
	\end{split}
	\end{align}
	together with
	\begin{align}
	\label{complex field transformations 2}
	\begin{split}
	\left[L_{-1}\,,O_{\Delta,J}(\x)\right]&=\partial_z O_{\Delta,J}(\x)\,,\\
	\left[L_0\,,O_{\Delta,J}(\x)\right]&=\frac{1}{2} \left(u\partial_u+2z\partial_z+2h \right)O_{\Delta,J}(\x)\,,\\
	\left[L_1\,,O_{\Delta,J}(\x) \right]&= z\left(u\partial_u+z\partial_z+2h \right)O_{\Delta,J}(\x)\,,
	\end{split}
	\end{align}
	and the conjugate relations. Here we defined the chiral weights 
	\begin{equation}
	\label{h hbar}
	h=\frac{\Delta+J}{2}\,, \qquad \bar h=\frac{\Delta-J}{2}\,.
	\end{equation}
	Note that one recovers the standard $SL(2,\mathbb{C})$ conformal field transformations by imposing $\partial_u O=0$, which ensures that the abelian translations $\tilde P_\mu$ act trivially in \eqref{complex field transformations}. Then $h,\bar h$ are the usual two-dimensional conformal weights.
	\begin{mytheo}{}{}
		Show that \eqref{infinitesimal coordinate change} is the infinitesimal version of the coordinate transformation $\x \mapsto \x'(\x)$:
		\begin{align}
		\label{complex coordinate transformations}
		\begin{split}
		z'&=z+a\,, \hspace{7cm} (\text{spatial translations})\\
		z'&=e^{i\theta} z\,, \hspace{8.1cm} (\text{rotation})\\
		z'&=\lambda z\,, \quad u'=\lambda u\,, \hspace{6.3cm} (\text{dilation})\\
		u'&=u+a^u\,, \hspace{7cm} (\text{time translation})\\
		u'&=u+b \zbar+\bar b z\,, \hspace{6.3cm} (\text{carroll boosts})\\
		z'&=\frac{z-k z\zbar}{1-k \zbar-\bar k z+k\bar k z\zbar}\,, \quad u'=\frac{u-k^u z\zbar}{1-k \zbar-\bar k z+k\bar k z\zbar}\,, \quad (\text{SCT})
		\end{split}
		\end{align}   
		and conjugate relations. Details may be found in \cite{Nguyen:2025sqk}.
	\end{mytheo}
	\begin{mytheo}{}{}
		Show that the transformations \eqref{complex field transformations}-\eqref{complex field transformations 2} are the infinitesimal version of
		\begin{equation}
		\label{conformal transfo}
		O'_{\Delta,J}(\x')=\left(\frac{\partial z'}{\partial z}\right)^{-h} \left(\frac{\partial \zbar'}{\partial \zbar}\right)^{-\bar h} O_{\Delta,J}(\x)\,.
		\end{equation}
	\end{mytheo}
	
	\paragraph{Indecomposable multiplets.} One can build carrollian field representations of $\operatorname{ISO}(1,3)$ that are more general than those presented in this section. In particular one can construct \textit{indecomposable representations}, described in terms of a carrollian primary field together with some partner field(s). An example is provided by the carrollian stress tensor multiplet $\{M(\x),N_i(\x)\}$. In absence of radiation, the mass aspect $M(\x)$ is a scalar primary of scaling dimension $\Delta=3$, while the transformation of $N_i(\x)$ is almost that of a spin-1 primary of scaling dimensions $\Delta=3$, except it gets an additional contribution proportional to the primary $M(\x)$. See \cite{Nguyen:2025sqk} and references therein. 
	
	\subsection{From particles to carrollian fields}
	We have looked at two seemingly distinct classes of $\operatorname{ISO}(1,3)$ representations: the massless particles of Wigner, known to be unitary and irreducible, and the carrollian conformal fields. The first are defined over the null momentum cone, while the latter are defined over $\scri$. 
	
	It turns out that they are essentially isomorphic to one another. To see this, all we need to do is to relate them via the following `Fourier-Mellin' transform  \cite{Banerjee:2018gce,Banerjee:2019prz,Bagchi:2022emh,Donnay:2022wvx,Nguyen:2023vfz}
	\begin{equation}
	\label{Mellin}
	O_{\Delta,J}(u,\vec x) =\int_0^\infty d\omega\, \omega^{\Delta-1} e^{i\omega u}\, a_J^\dagger(p(\omega,\vec x))\,,    
	\end{equation}
	where $a^\dagger_J(p(\omega,\vec x))$ is a creation operator for a massless particle of helicity $J$ and momentum $p(\omega,\vec x)$ as defined in \eqref{massless momentum parametrisation}.
	This provides a simple intertwining relation between the particle states of a unitary theory and the carrollian conformal fields. It also shows that $\scri$ is the Fourier conjugate of the null momentum cone.
	
	\begin{mytheo}{}{}
		Check consistency of the intertwining relation \eqref{Mellin} by acting with $\operatorname{ISO}(1,3)$ generators on both sides of the equation.   
	\end{mytheo}
	
	There is one point left to discuss though: what is the value of $\Delta$? Up to now the scaling dimension $\Delta$ has been left totally unconstrained, and does not seem to relate to any quantum number of the momentum particles. A natural choice would be $\Delta=1$ such that \eqref{Mellin} reduces to a simple Fourier transform. We will see in section~\ref{sec: holographic dictionary} that this value arises naturally from the `holographic dictionary', i.e., by studying the correspondence with bulk fields in asymptotically flat spacetimes. But there is at least one reason to keep $\Delta$ a free parameter in \eqref{Mellin}, which is that we can shift its value by taking $u$-derivatives:
	\begin{equation}
	\label{Mellin derivatives}
	(-i\partial_u)^n\, O_{\Delta,J}(u,\vec x) =\int_0^\infty d\omega\, \omega^{\Delta+n-1} e^{i\omega u}\, a_J^\dagger(p(\omega,\vec x))\,,   \qquad n \in \mathbb{N}\,.
	\end{equation}
	Thus even if we imagine dealing with fundamental primary fields of scaling dimension $\Delta=1$, we can always construct primary fields of scaling dimension $\Delta=1+n$ by taking $n$ derivatives. 
	
	\subsection{From scattering amplitudes to carrollian correlators}
	\label{subsec:bootstrap}
	Given the correspondence \eqref{Mellin} between massless particles and carrollian conformal fields, it is natural to speculate that massless scattering amplitudes map to carrollian correlation functions via
	\begin{equation}
	\label{Mellin amplitudes}
	\langle O^{\eta_1}_{\Delta_1,J_1}(\x_1)\, ...\, O^{\eta_n}_{\Delta_n,J_n}(\x_n) \rangle\equiv \prod_{k=1}^n \int_0^\infty d\omega_k\, \omega^{\Delta_k-1} e^{i \eta_k \omega_k u_k} S_n(1^{J_1}...\,n^{J_n})\,,
	\end{equation}
	where $\eta_k=\pm 1$ depending whether the particle is ingoing (+) or outgoing (-), with momenta parametrized as
	\begin{equation}
	\label{pk}
	p_k^\mu=\eta_k\, \omega_k\, q^\mu(\vec x_k)\,,
	\end{equation}
	where $q^\mu(\vec x)$ was defined in \eqref{massless momentum parametrisation}.
	This is a standard convention where all particles can be effectively treated as if they were ingoing, with ingoing momenta $p_k^\mu$ as given above and ingoing helicity $J_k$ (physical helicity is $\eta_k J_k) $. The in/out label $\eta$ is a `flavor' index referring to whether the corresponding carrollian field should be thought of as living on the future or past component of null infinity, although we will often drop it for notational convenience. Because the $S$-matrix elements $S_n(1^{J_1}...\,n^{J_n})$ on the right-hand side of \eqref{Mellin amplitudes} transform like products of massless particles, the left-hand side necessarily transforms like products of carrollian fields. The objects defined in \eqref{Mellin amplitudes} are called \textit{carrollian amplitudes}. Now we come to the questions which lie at the core of these lectures on Carrollian Holography:
	\begin{quote}
		\begin{center}
			\emph{Can we compute carrollian amplitudes independently of Feynman diagrams?}\\
			\emph{Can we compute them non-perturbatively using CFT-like methods?}
		\end{center}
	\end{quote}
	To answer these questions, we will have to build a general theory of carrollian correlators. We will follow a \textit{bootstrap formulation}, i.e., we will try to complete the following program:
	\begin{program*}{}{}
		\begin{itemize}
			\item[1.] Classify states and fields that are potentially relevant to the (gravitational) scattering theory we are interested in,
			\item[2.] Classify kinematically allowed 2-, 3-, and 4-point correlation functions for these fields,
			\item[3.] Establish the existence of a convergent operator product expansion (OPE),
			\item[4.] Determine the `carrollian conformal blocks' that form a basis of kinematically allowed 4-point functions,
			\item[5.] Determine consistency conditions, such as crossing equations, which follow from the previous steps.   
		\end{itemize}
	\end{program*}
	Such a bootstrap formulation has been extremely successful for conventional CFTs, i.e., conformal theories with unitary $\operatorname{SO}(2,d)$ symmetry, as it allows to identify small isolated islands in parameter space where a consistent CFT may be found \cite{Poland:2018epd,Rychkov:2023wsd}. Reaching a similar level of predictivity with carrollian amplitudes would amount to a novel bootstrap formulation of massless scattering theory, one ideally suited to non-perturbative gravitational scattering.   
	
	In this section, we have already achieved step 1 of this program to a large extent. We will tackle step 2 and 3 in Sections~\ref{sec: carrollian correlators and amplitudes} and \ref{sec:OPE}, respectively. Steps 4 and 5 lie beyond the current state-of-the-art, although the field is rapidly evolving. But for now, let us give the first piece of evidence that the whole program may actually work, by looking at the simplest set of observables: 1-to-1 scattering amplitudes and their relation to 2-point carrollian correlators. 
	
	\subsubsection*{Carrollian 2-point functions} 
	\label{subsec: carrollian 2-pt functions}
	Let us look at the kinematic constraints on the correlation functions of generic carrollian conformal primary fields of the type described in Section~\ref{subsec: carrollian fields}. As with any other symmetry, these constraints come in the form of Ward identities, 
	\begin{equation}
	\label{Ward identities}
	\sum_{i=1}^n \langle O_1(\x_1)\,...\,\delta O_i(\x_i)\,...\, O_n(\x_n) \rangle=0\,,
	\end{equation}
	where $\delta O_i$ is any linear combination of the infinitesimal $\operatorname{ISO}(1,3)$ transformations given in \eqref{Carrollian induced rep}. In particular, translation invariance generated by $P_\alpha=(H,P_i)$ implies that correlation functions only depend on the insertion points through the separations $\x_{ij}\equiv \x_i-\x_j$. As a set of independent variables we can therefore choose $\x_{i\, i+1}$ with $i=1,...,n-1$. 
	
	For a two-point function
	\begin{equation}
	C_2(\x_{12})\equiv \langle O_1(\x_1)O_2(\x_2) \rangle\,,
	\end{equation}
	the Ward identity \eqref{Ward identities} associated with the `carroll boost' $B_i$ can be written
	\begin{equation}
	x_{12}^i\, \partial_{u_{12}} C_2(\x_{12})=0\,,
	\end{equation}
	whose distributional solutions are of the form
	\begin{equation}
	\label{C2 solution}
	C_{12}(\x_{12})=f_{12}(\vec x_{12})+  g_{12}(u_{12})\, \delta(\vec x_{12})\,.
	\end{equation}
	Without loss of generality we can assume that all contact terms are captured by the second term, such that $f(\vec x_{12})$ should independently solve the Ward identities \eqref{Ward identities}. Looking at \eqref{Carrollian induced rep} we see that for this term they reduce to the conformal Ward identities in $\mathbb{R}^2$, with well-known solutions\begin{equation}
	\label{CFT 2-point function}
	f_{12}(\vec x)=\# \frac{\delta_{\Delta_1,\Delta_2}\, \delta_{J_1,J_2}}{|\vec x|^{\Delta_1+\Delta_2}}\,.
	\end{equation}
	We then turn to the determination of the other branch of solutions given in terms of $g_{12}(u)$. Rotation invariance imposes the helicity of the two fields to be conjugate of each other, and we can therefore write
	\begin{equation}
	\label{C2 solution bis}
	g_{12}(u)=\delta_{J_1,-J_2}\, \tilde g_{12}(u)\,.
	\end{equation}
	\begin{mytheo}{}{}
		Show that the dilation Ward identity implies
		\begin{equation}
		\left(\Delta_1+\Delta_2+u\partial_u-2\right)\tilde g_{12}(u)=0\,,
		\end{equation}
		with solution
		\begin{equation}
		\tilde g_{12}(u)=\begin{cases}
		\#\, u^{2-\Delta_1-\Delta_2} \qquad &\text{for} \quad \Delta_1+\Delta_2 \neq 2\,,\\
		\#+\#\, \sgn (u) \qquad &\text{for} \quad \Delta_1+\Delta_2 =2\,.
		\end{cases}\
		\end{equation}
	\end{mytheo}
	\noindent One can check that invariance under special conformal transformations generated by $K_\alpha=(K,K_i)$ do not impose extra restrictions. More details can be found in \cite{Donnay:2022wvx,Nguyen:2023miw,Nguyen:2025sqk}.
	
	In summary, for generic scaling dimensions the kinematically allowed 2-point carrollian correlators are of the form
	\begin{equation}
	\label{2-pt functions}
	\langle O_{\Delta,J_1}(\x_1) O_{\Delta_2,J_2}(\x_2) \rangle=a\, \frac{ \delta_{\Delta_1,\Delta_2}\, \delta_{J_1,J_2}}{|\vec x_{12}|^{\Delta_1+\Delta_2}}+b\, \frac{ \delta_{J_1,-J_2}\, \delta(\vec x_{12})}{(u_{12})^{\Delta_1+\Delta_2-2}}\,.
	\end{equation}
	The first term is identical to the 2-point function in conventional CFT$_2$. The second term is a contact term in space, but it allows non-trivial time-dependence. Two-point functions are therefore less constrained by kinematics in carrollian CFTs than they are in conventional CFTs. This will be a generic feature of carrollian CFT.

	\subsubsection*{Massless 2-point amplitudes}
	Let us now perform a simple test of the correspondence between massless scattering amplitudes and carrollian conformal correlators. We start with the unique 1-to-1 scattering amplitude, equal to the Lorentz-invariant inner product
	\begin{equation}
	S_2(1^{J_1}2^{J_2})=(\omega_1)^{-1} \delta(\omega_1-\omega_2) \delta(\vec x_{12})\delta_{J_1,-J_2}\,, \qquad \eta_2=1=-\eta_1\,.
	\end{equation}
	Application of \eqref{Mellin amplitudes} yields the carrollian amplitude
	\begin{equation}
	\label{divergent integral}
	\begin{split}
	\langle O^+_{\Delta_1,J_1}(\x_1)\, O^-_{\Delta_2,J_2}(\x_2) \rangle&=\int_0^\infty \dd \omega_1 \dd \omega_2\, \omega_1^{\Delta_1-1} \omega_2^{\Delta_2-1} e^{-i\omega_1 u_1}e^{i\omega_2 u_2}\, S_2(1^{J_1}2^{J_2})\\
	&=\delta(\vec x_{12})\delta_{J_1,-J_2} \int_0^\infty \dd \omega\,  \omega^{\Delta_1+\Delta_2-3}\, e^{-i\omega u_{12}}\,.
	\end{split}
	\end{equation}
	For convergence of the above integral, we need to give a small negative imaginary part to $u_{12}$, and we obtain \cite{Banerjee:2018gce,Liu:2022mne,Bagchi:2022emh,Donnay:2022wvx}
	\begin{equation}
	\label{free transformed amplitude}
	\langle O^+_{\Delta_1,J_1}(\x_1) O^-_{\Delta_2,J_2}(\x_2) \rangle=\Gamma[\Delta_1+\Delta_2-2]\, \frac{\delta(\vec x_{12})\delta_{J_1,-J_2}}{(iu_{12})^{\Delta_1+\Delta_2-2}}\,.
	\end{equation}
	We observe that this indeed takes the form of a carrollian conformal correlator \eqref{2-pt functions}, provided $\Delta_1+\Delta_2 \neq 2$. This is the carrollian two-point function that we should therefore select in the scattering context. 
	
	For the special value $\Delta_1+\Delta_2 = 2$, the integral \eqref{divergent integral} diverges from its lower bound, which is also seen in \eqref{free transformed amplitude} from the corresponding pole in $\Gamma[\Delta_1+\Delta_2-2]$.  The carrollian amplitude \eqref{free transformed amplitude} can be expanded in $\beta\equiv \Delta_1+\Delta_2-2$ \cite{Liu:2022mne,Donnay:2022wvx},
	\begin{equation}
	\label{beta divergences}
	\langle O^+_{\Delta_1,J_1}(\x_1) O^-_{\Delta_2,J_2}(\x_2) \rangle=\left(\frac{1}{\beta}-\gamma_E-\ln (iu_{12})+O(\beta^2) \right) \delta(\vec x_{12})\,.
	\end{equation}
	This is not a valid carrollian 2-pt function (unless we keep all terms in the $\beta$ expansion), so we should stick to $\Delta_1+\Delta_2\neq 2$. Even though we have argued that $\Delta=1$ may be a natural choice of scaling dimension, in practice we have to consider more general values, for instance those obtained by taking time-derivatives as in \eqref{Mellin derivatives}. This prescription will be justified further in the next section.
	
	\section{Holographic dictionary}
	\label{sec: holographic dictionary}
	In the previous section, we made a direct connection between the null momentum cone and null infinity $\scri$, where massless particle states and carrollian conformal fields are resectively naturally defined. This correspondence does not rely in any way on Minkowski spacetime $\mathbb{M}^4$ or any other asymptotically flat spacetime, even though the latter played an important role in defining $\scri$ in Section~\ref{sec: AFS}. This is indeed the whole purpose of Carrollian Holography: to go beyond classical spacetimes and enter the realm of non-perturbative quantum gravity.
	
	Of course nothing prevents us from going back to perturbative QFT in $\mathbb{M}^4$, perhaps as a way to strengthen the intuition we are currently developing. This is what we will briefly do in this section. More specifically, we will show that the pullback to $\scri$ of the massless relativistic quantum fields in $\mathbb{M}^4$ provides realizations of the carrollian conformal fields introduced in Section~\ref{subsec: carrollian fields}. This `extrapolate dictionary' will be shown to work also at the level of time-ordered correlators.
	
	\subsection{Extrapolation of fields}
	Thus we return to Minkowski spacetime $\mathbb{M}^4$ in cartesian coordinates $X^\mu$. We introduce a set of retarded coordinates $(r,u,\vec x)$ through
	\begin{equation}
	\label{coord transf retarded}
	X^\mu=u\, n^\mu+r\, q^\mu(\vec x)\,, \qquad  n^\mu\equiv \frac{1}{\sqrt{2}}(1,\vec 0,-1)\,,
	\end{equation}
	where $q^\mu(\vec x)$ is again given by \eqref{massless momentum parametrisation}. 
	\begin{mytheo}{}{}
		Show that $n^\mu$ and $q^\mu(\vec x)$ are null vectors, and that they satisfy the properties
		\begin{equation}
		n \cdot q=-1\,, \qquad q(\vec x) \cdot q(\vec y)=-|\vec x-\vec y|^2\,.
		\end{equation}
	\end{mytheo}
	\noindent In these coordinates, the Minkowski metric takes the simple form
	\begin{equation}
	\label{flat Bondi gauge}
	d \tilde s^2=\eta_{\mu\nu} \dd X^\mu \dd X^\nu =-2 \dd u \dd r+ 2 r^2 \delta_{ij} \dd x^i \dd x^j\,,
	\end{equation}
	which lies in the class of Bondi metrics \eqref{Bondi expansion} with flat boundary representative $q_{ij}=\delta_{ij}$. Hence,
	future null infinity $\scri^+$ is locates at $r \to \infty$. In fact, this coordinate system has the advantage that past null infinity $\scri^-$ is reached in the opposite limit $r \to -\infty$. A detailed discussion of this coordinate system may be found in appendix A of \cite{Donnay:2022wvx}.
	
	Let us then consider the photon field $A_\mu$ as a concrete example, which may be easily generalized to any massless field \cite{Donnay:2022wvx,Nguyen:2023vfz}. In the free theory, it admits an expansion into particle operators,
	\begin{equation}
	A_\mu(X)=\sum_{i=1,2} \int [d^3p]\, \varepsilon_\mu^i(p)\, e^{-i p\cdot X}\, a^\dagger_i (p)+ \text{h.c.}\,,
	\end{equation}
	where the sum runs over two independent photon polarizations. Let us again adopt the convenient parametrization $p(\omega,\vec y)=\omega q(\vec y)$ as given in \eqref{massless momentum parametrisation}. A suitable choice of polarization vector is then given by
	\begin{equation}
	\varepsilon^\mu_i(p)=\frac{1}{\sqrt{2}}\frac{\partial q^\mu(\vec y)}{\partial y^i}=\varepsilon^\mu_i(\vec y)\,.
	\end{equation}
	\begin{mytheo}{}{}
		Show that the polarization vectors satisfy
		\begin{equation}
		p_\mu \varepsilon^\mu_i(p) =0\,, \qquad \varepsilon^\mu_j(p) \varepsilon_\mu^i(p)=\delta^i_j.
		\end{equation}
		Show that $\varepsilon^\mu_i(\vec x)$ is a projector onto the space tangent to the celestial sphere.
	\end{mytheo}
	\noindent The photon field can thus be written
	\begin{equation}
	A_\mu(r,\x)=\sum_{i=1,2} \int_0^\infty d \omega\, \omega \int d^2\vec y\,  \varepsilon_\mu^i(\vec y)\, e^{i\omega (u+r|\vec x-\vec y|^2)}\, a^\dagger_i (\omega,\vec y)+ \text{h.c.}\,,
	\end{equation}
	Now we want to pull this field back to $\scri^+$, i.e., we want to evaluate it in the limit $r \to \infty$. The $\vec y$-integral can be performed by stationary phase approximation, which localizes the integrand at $\vec y=\vec x$ and yields
	\begin{equation}
	A_\mu \approx -\frac{i\pi}{r} \sum_{i=1,2} \int_0^\infty \dd\omega\, \varepsilon_\mu^i(\vec x)\, e^{i\omega u}\, a^\dagger_i(\omega,\vec x)+\text{h.c.}
	\end{equation}
	Thus, we can define
	\begin{equation}
	\label{extrapolate field}
	O_i(\x)\equiv \lim_{r \to \infty} \frac{ir}{\pi}\, \varepsilon_i^\mu(\vec x) A_\mu(r,\x)= \int_0^\infty \dd\omega\, e^{i\omega u}\, a^\dagger_i(\omega,\vec x)+\text{h.c.}
	\end{equation}
	This is a spin-1 carrollian primary field of scaling dimension $\Delta=1$. The only difference with the formula \eqref{Mellin} is that we are not using the helicity basis. Rather, the spin-1 states are encoded in the vector representation of $\operatorname{SO}(2)$, but this is of course completely equivalent. In addition, \eqref{extrapolate field} is hermitian as it also contains the annihilation part. 
	
	This holographic correspondence indicates that $\Delta=1$ is perhaps the natural choice of scaling dimension. However, we have seen in \eqref{beta divergences} that this value yields divergent carrollian amplitudes, and we have argued that we should take at least one $u$-derivative on one of the carrollian primary field, thereby shifting its dimension by one unit. This prescription actually has a natural justification from the perspective of bulk field reconstruction. Indeed, one can reconstruct $A_\mu(X)$ at any point $X \in \mathbb{M}^4$ from the knowledge of $\partial_u O_i(\x)$ rather than $O_i(\x)$ itself. See the Kirchhoff-d’Adh\'emar formula in \cite{Donnay:2022wvx}.
	
	This `extrapolate dictionary' between bulk and boundary fields can be extended to arbitrary integer spin and spacetime dimension $\mathbb{M}^{d+1}$. It reads \cite{Nguyen:2023vfz}
	\begin{equation}
	\label{extrapolate field: general case}
	\begin{split}
	O_{i_1...\,i_s}(\x)&\equiv \lim_{r \to \infty} \left(\frac{ir}{\pi}\right)^{\frac{d-1}{2}}\, \varepsilon_{i_1}^{\mu_1}(\vec x)...\, \varepsilon_{i_s}^{\mu_s}(\vec x) \phi_{\mu_1...\,\mu_s}(r,\x)\\
	&= \int_0^\infty \dd\omega\, \omega^{\frac{d-3}{2}} e^{i\omega u}\, a^\dagger_{i_1...\,i_s}(\omega,\vec x)+\text{h.c.}\,,
	\end{split}
	\end{equation}
	where $\vec x \in \mathbb{R}^{d-1}$ is now a stereographic coordinate on the celestial sphere $S^{d-1}$. This is a spin-$s$ carrollian primary field of scaling dimension $\Delta=\frac{d-1}{2}$.
	By contrast to massless fields in $\mathbb{M}^{d+1}$, the boundary carrollian fields at $\scri_d$ are not constrained by a wave equation or any other analogue. The group theoretical reason for this is that the Casimir constraint $\mathcal{C}_2=0$ is automatically satisfied in the carrollian primary field representation, as anticipated in \eqref{C2=0}. Another way to look at it is that carrollian conformal fields contain just the right number of degrees of freedom to encode the massless particle states. Indeed, $\scri$ is Fourier conjugate to the null momentum cone such that carrollian fields are automatically on-shell. The spin states organize themselves into symmetric-traceless tensor representations of the massless little group $\operatorname{SO}(d-1)$, and the corresponding tensor indices are identified with frame indices tangent to the celestial sphere.  
	
	\subsection{Extrapolation of correlators}
	Given the holographic dictionary for fields given in \eqref{extrapolate field: general case}, it is natural to introduce a similar notion of \textit{holographic carrollian correlator} \cite{Nguyen:2023miw},
	\begin{equation}
	\label{procedure}
	\langle O_1^{\eta_1}(\x_1)\,...\, O_n^{\eta_n}(\x_n) \rangle\equiv \mathcal{N} \lim_{r \to \infty} r^{\frac{n(d-1)}{2}}\, \langle \mathcal{T}\{ \phi_1(\eta_1 r,\x_1)\,...\, \phi_n(\eta_n r,\x_n) \}\rangle\,,
	\end{equation}
	where spin indices on the left, and contraction with the corresponding polarization tensors on the right are left implicit, and where $\mathcal{N}$ is a normalization of our choice. Here we chose to focus on time-ordered correlators, but Wightman, retarded/advanced correlators may also be considered \cite{Nguyen:2023miw}. We expect this quantity to precisely take the form of a correlator for the corresponding carrollian conformal fields. Although this definition is distinct from that of carrollian amplitudes given in \eqref{Mellin}, we will see that they are in fact closely related. 
	
	\paragraph{Scalar 2-point function.}
	We start with the simple example of a two-point function for a massless scalar, which will already illustrate interesting features of the correspondence between time-ordered correlators in $\mathbb{M}^4$ and holographic carrollian correlators as defined through \eqref{procedure}. First we recall that the `dual' pair of fields in this case is simply
	\begin{equation}
	\phi(X) \quad \longleftrightarrow \quad O_\Delta(\x)\,,
	\end{equation}
	with canonical scaling dimension $\Delta=1$. Indeed both types of fields encode massless scalar particles. The time-ordered two-point function, or Feynman propagator, is given by 
	\begin{equation}
	\label{Feynman propagator}
	\langle \mathcal{T}\{\phi(X_1) \phi(X_2)\}\rangle=\frac{1}{\lambda + i 0^+}=\frac{\lambda}{\lambda^2+0^+}-i\pi \delta(\lambda)\,,
	\end{equation}
	with $\lambda=(X_1-X_2)^2$ the invariant square distance. This distribution is split between two contributions: the second term in \eqref{Feynman propagator} encodes the lightcone `singularity', while the first term -- known as Cauchy's principal value distribution -- interpolates between $\lambda^{-1}$ at $\lambda\neq 0$ and zero at $\lambda=0$. Now we adopt retarded coordinates using \eqref{coord transf retarded}, such that
	\begin{equation}
	\label{lambda}
	\lambda = (X_1-X_2)^2=2r_1r_2 |x_{12}|^2-2u_{12} r_{12}\,.
	\end{equation}
	We change variables $\lambda \mapsto |x_{12}|^2$ using \eqref{lambda} and evaluate these distributions in the limit $r\equiv r_1=-r_2\to \infty$. This will allow us to evaluate the \textit{in-out} holographic correlator
	\begin{equation}
	\label{out out procedure}
	\langle O_{\Delta}^+(\x_1) O^-_{\Delta}(\x_2) \rangle\equiv \lim_{r \to \infty} (-2r^2)\, \langle \mathcal{T}\{\phi(r,\x_1) \phi(-r,\x_2)\}\rangle\,.
	\end{equation}
	For the delta distribution we simply get
	\begin{equation}
	\delta(\lambda)=\frac{1}{2r^2}\, \delta(|\vec x_{12}|^2)=\frac{\pi}{2r^2}\, \delta(\vec x_{12})\,.
	\end{equation}
	For the first term in \eqref{Feynman propagator}, we introduce the variable $\rho^2=r/(2 u_{12})$ such that we can write
	\begin{equation}
	\frac{\lambda}{\lambda^2+0^+}=-\frac{1}{2r^2} \frac{\rho^{2}}{1+\rho^2|x_{12}|^2}\,.
	\end{equation}
	The limit $\rho \to \infty$ of this distribution is explicitly given by \cite{Donnay:2022ijr}
	\begin{equation}
	\frac{\rho^{2}}{1+\rho^2|x_{12}|^2}=\pi\, \ln \rho^2\, \delta(\vec x_{12})+\frac{1}{|x_{12}|^{2}}+...\,,
	\end{equation}
	Altogether, we thus find
	\begin{equation}
	\label{divergent 2-pt function}
	\langle O_{\Delta}^+(\x_1) O^-_{\Delta}(\x_2) \rangle=\pi \left(\ln \frac{r}{2}-\ln u_{12}\right) \delta(\vec x_{12})+\frac{1}{|x_{12}|^{2}}+ i\pi^2 \,  \delta(\vec x_{12})\,.
	\end{equation}
	We observe that the last two terms are kinematically allowed carrollian two-point functions \eqref{2-pt functions} for scalar operators of scaling dimension $\Delta=1$. The first term is strictly divergent as it contains $\ln r$ in the limit $r\to \infty$. Moreover, it does not display the expected $u$-dependence. But closer inspection reveals that this contribution is of the exact same form as \eqref{beta divergences} upon identifying $\ln r \sim \beta^{-1}$. We have thus re-discovered this infrared divergence from a bulk perspective! By now, we know how to deal with this issue: we need to take at least one $u$-derivative. This yields for instance
	\begin{equation}
	\langle O_{\Delta}^+(\x_1) \partial_{u_2}  O^-_{\Delta}(\x_2) \rangle=\frac{\pi \delta(\vec x_{12})}{u_{12}}\,.
	\end{equation}
	This is a well-defined carrollian two-point function for operators of scaling dimensions $\Delta_1=1$ and $\Delta_2=2$. Moreover, it agrees with the corresponding carrollian amplitude \eqref{free transformed amplitude}!
	
	Of course we knew that there is a relation between time-ordered correlators and amplitudes, but this would normally involve the LSZ reduction procedure. The role of the LSZ formula is to put external legs on the mass shell. In the `holographic' approach taken here, this is automatically implemented by pulling back insertion points to $\scri$, which is the Fourier conjugate of the null momentum shell.
	
	\paragraph{Photon and graviton 2-point functions.}
	We now apply the same procedure to the photon and graviton propagators. Since they are constructed from the scalar propagators, they will inherit the features described above. We start from the propagators in momentum space in a generic $\xi$-gauge. For the photon propagator we use the standard formula
	\begin{equation}
	G_{\mu\nu}(p)=\left(\eta_{\mu\nu}-(1-\xi)\, \frac{p_\mu p_\nu}{p^2} \right) \frac{1}{p^2}\,,
	\end{equation}
	while the corresponding expression for the graviton propagator is given by \cite{Jakobsen:2020diz,Prinz:2020nru}
	\begin{equation}
	G_{\alpha\beta}^{\mu\nu}(p)=\left( P_{\alpha \beta}^{\mu\nu}-\frac{1}{2} \eta_{\alpha \beta}\, \eta^{\mu\nu}-2(1-\xi) P^{\mu\nu}_{\rho\kappa}\, \frac{p^\rho p_\sigma}{p^2} P^{\kappa\sigma}_{\alpha \beta}   \right)\frac{1}{p^2}\,,
	\end{equation}
	in terms of the projector onto the space of symmetric tensors
	\begin{equation}
	P^{\mu\nu}_{\alpha\beta}=\frac{1}{2}\left(\delta^\mu_\alpha \delta^\nu_\beta+\delta^\mu_\beta \delta^\nu_\alpha\right)\,.
	\end{equation}
	After Fourier transforming back to position space we obtain the bulk two-point functions
	\begin{align}
	\langle A_\mu(X_1) A_\nu(X_2) \rangle&=\left(\eta_{\mu\nu}-(1-\xi)\, \frac{X^{12}_\mu\, X^{12}_\nu}{(X_{12})^2}  \right) G_F(\lambda)\,,\\
	\langle h^{\mu\nu}(X_1)\, h_{\alpha\beta}(X_2) \rangle&=\left( P_{\alpha \beta}^{\mu\nu}-\frac{1}{d-1} \eta_{\alpha \beta}\, \eta^{\mu\nu}-2(1-\xi) P^{\mu\nu}_{\rho\kappa}\, \frac{X_{12}^\rho\, X^{12}_\sigma}{(X_{12})^2} P^{\kappa\sigma}_{\alpha \beta}   \right) G_F(\lambda)\,,
	\end{align}
	where $G_F(\lambda)$ is the scalar propagator given in \eqref{Feynman propagator}. 
	\begin{mytheo}{}{}
		In retarded coordinates, show that
		\begin{align}
		\langle A_i(r,\x_1) A_j(-r,\x_2) \rangle&=2\left(\delta_{ij}-(1-\xi)\, \frac{x^{12}_i\, x^{12}_j}{|x_{12}|^2}  \right) G_F(\lambda)\,,\\
		\langle h^{ij}(r,\x_1) h_{kl}(-r,\x_2) \rangle&=4\left( P^{ij}_{kl}-2(1-\xi) P^{ij}_{mn}\, \frac{x_{12}^m\, x^{12}_p}{|x_{12}|^2} P^{np}_{kl}   \right) G_F(\lambda)\,.
		\end{align}
	\end{mytheo}
	At this point we notice that the gauge choice $\xi=-1$ is such that the expressions in brackets reduce to the inversion tensors familiar from conformal field theory in $\mathbb{R}^{2}$. Indeed for $\xi=-1$ we have
	\begin{equation}
	\begin{split}
	\langle A_i(r,\x_1) A_j(-r,\x_2) \rangle&=2\,\mathcal{I}_{ij}(x_{12}) G_F(\lambda)\,,\\
	\langle h_{ij}(r,\x_1)\, h_{kl}(-r,\x_2) \rangle&=4\,\mathcal{I}_{im}(x_{12}) \mathcal{I}_{jn}(x_{12}) \mathcal{P}^{mn}_{kl}\, G_F(\lambda)\,,
	\end{split}
	\end{equation}
	with
	\begin{equation}
	\mathcal{I}_{ij}(\vec x)=\delta_{ij}-\frac{2x_i x_j}{|\vec x|^2}\,, \qquad \mathcal{P}^{mn}_{ij}=\frac{1}{2}\left(\delta^m_i \delta^n_j+\delta^m_j \delta^n_i\right)-\frac{1}{2}\, \delta^{mn} \delta_{ij}\,.
	\end{equation}
	Finally, we apply the definition \eqref{procedure} in order to obtain the corresponding holographic carrollian correlators,
	\begin{equation}
	\begin{split}
	\langle O^+_i(\x_1) O^-_j(\x_2) \rangle&=\mathcal{I}_{ij}(\vec x_{12}) \langle O^+(\x_1) O^-(\x_2) \rangle\,,\\
	\langle O^+_{ij}(\x_1) O^-_{kl}(\x_2) \rangle&=\mathcal{I}_{im}(\vec x_{12}) \mathcal{I}_{jn}(\vec x_{12}) \mathcal{P}^{mn}_{kl}\, \langle O^+(\x_1) O^-(\x_2) \rangle\,,
	\end{split}
	\end{equation}
	in terms of the scalar correlator \eqref{divergent 2-pt function}. Because the latter is divergent, we again need to take at least one $u$-derivative. The resulting 2-point functions are well-defined carrollian correlators for spin-1 and spin-2 primary fields. See \cite{Nguyen:2023miw} for more details and generalization to arbitrary dimension. See also \cite{Kulkarni:2025qcx} for discussions involving higher point functions. 
	
	In the rest of these lectures, we will not investigate further this holographic dictionary but rather focus our attention on the physical observables: amplitudes and correlators. 
	
	\section{Carrollian correlators and amplitudes}
	\label{sec: carrollian correlators and amplitudes}
	We now return to the development of a carrollian conformal field theory which could be used to determine massless scattering amplitudes. In this section we will carry out step 2 of the bootstrap program set out in Section~\ref{subsec:bootstrap}, i.e., we will classify 2-, 3-, and 4-point functions allowed by kinematics. We will then assess which of these correspond to massless scattering amplitudes.
	
	The first point to discuss is the need to complexify the particle's momenta. For this, let us consider the scattering of three massless particles.
	\begin{mytheo}{}{}
		Show that momentum conservation $p_1+p_2+p_3=0$, together with $p_i^2=0$, implies 
		\begin{equation}
		\label{colinearity}
		p_1 \cdot p_2=p_2 \cdot p_3=p_1 \cdot p_3=0\,.
		\end{equation}
		In the parametrization \eqref{massless momentum parametrisation}, this reads
		\begin{equation}
		\label{colinearity 2}
		|\vec x_{12}|^2=|\vec x_{13}|^2=|\vec x_{23}|^2=0\,.
		\end{equation}
	\end{mytheo}
	\noindent Indeed, it is well-known that momentum conservation implies that a set of three massless particles can interact only if they are strictly colinear. Thus for real momenta, the support of the corresponding 3-point scattering amplitude vanishes. To evade this problem, we will use a common trick, which is to adopt the complex stereographic coordinates $(z,\zbar)$ defined in \eqref{z zbar} and consider them independent variables. Real kinematics can be recovered by imposing the reality condition $\zbar=z^*$. The condition \eqref{colinearity} is then satisfied by
	\begin{equation}
	\label{complex colinearity}
	z_{12}=z_{13}=z_{23}=0\,, \qquad \text{or} \qquad \zbar_{12}=\zbar_{13}=\zbar_{23}=0\,,
	\end{equation}
	which are weaker than \eqref{colinearity 2} provided $z,\zbar$ are independent. This allows to define 3-point massless amplitudes that are non-trivial as complex distributions, which are now commonly used in on-shell constructions of massless scattering amplitudes \cite{Elvang:2015rqa}. We will give their expression in Section~\ref{subsec: carrollian amplitudes}.
	
	\subsection{Carrollian correlators of complex kinematics}
	In this section we present a list of 2-, 3-, 4-point carrollian correlators that solve the $\operatorname{ISO}(1,3)$ Ward identities \eqref{Ward identities}. We saw in Section~\ref{subsec: carrollian 2-pt functions} that 2-point carrollian correlators with real kinematics admit two distinct classes of solutions. Only one is related to unitary scattering theory, however. The same situation arises for higher-point functions \cite{Nguyen:2023miw,Nguyen:2025sqk}. For brevity, we will only discuss classes of solutions related to scattering amplitudes.
	
	\paragraph{2-point functions.}
	We have already seen in Section~\ref{subsec:bootstrap} that the relevant class of 2-point carrollian correlators is given by
	\begin{equation}
	\langle O^+_{\Delta_1,J_1}(\x_1) O^-_{\Delta_2,J_2}(\x_2) \rangle=\# \frac{\delta(\vec x_{12})\delta_{J_1,-J_2}}{(u_{12})^{\Delta_1+\Delta_2-2}}\,.
	\end{equation} 
	
	\paragraph{3-point functions.}
	The first step is to determine the quantities constructed out of the three coordinates $\x_1,\x_2,\x_3$ which are translation-invariant and transform covariantly under \eqref{complex coordinate transformations}. In general, only the separations $\x_{12}$ have this property, however on the support $\zbar_1=\zbar_2=\zbar_3\equiv \zbar$ discussed in \eqref{complex colinearity}, we can also consider the quantity
	\begin{equation}
	F_{123}\equiv u_1 z_{23}+u_2 z_{31}+u_3 z_{12}\,.
	\end{equation}
	\begin{mytheo}{}{}
		Show that $F_{123}$ transforms as    
		\begin{equation}
		\begin{aligned}
		F_{123}'&=e^{i\theta} F_{123}\,, &\qquad &(\text{rotation})\,,\\
		F_{123}'&=\lambda^2 F_{123}\,, &\qquad &(\text{dilation})\,,\\
		F_{123}'&=\frac{F_{123}}{1-k \zbar}\,,  &\qquad &(\text{SCT})\,,\\
		F_{123}'&=\frac{F_{123}}{(1-\bar k z_1)(1-\bar k z_2)(1-\bar k z_3)}\,, &\qquad &(\text{SCT})\,,
		\end{aligned}
		\end{equation}
		while it is invariant under all remaining Poincar\'e symmetries.
	\end{mytheo}
	\noindent Using this we are able to write `chiral' 3-point functions supported on $\zbar_1=\zbar_2=\zbar_3$, which transform like products of fields following \eqref{conformal transfo}. Translation and carroll boost invariance imply that the chiral 3-point function is a function of the coordinates through $z_{ij}$ and $F_{123}$, while covariance under conformal transformations fixes its form to be
	\begin{equation}
	\label{chiral 3-point function}
	\langle O_1 O_2 O_3 \rangle=\# \frac{\delta(\zbar_{12})\delta(\zbar_{23})}{ (z_{12})^a\, (z_{23})^b\, (z_{13})^c\, (F_{123})^d}\,.
	\end{equation}
	\begin{mytheo}{}{}
		Show that dilation, rotation, and special conformal transformations imply    
		\begin{equation}
		\label{abcd}
		\begin{split}
		a=J_1+J_2-\Delta_3+2\,, \qquad b&=J_2+J_3-\Delta_1+2\,, \qquad c=J_1+J_3-\Delta_2+2\,,\\
		d&=2\bar h_1+2\bar h_2+2\bar h_3-4\,,
		\end{split}
		\end{equation}
		with $h,\bar h$ defined in \eqref{h hbar}.
	\end{mytheo}
	\noindent Of course, one can write down the `anti-chiral' solution, obtained from \eqref{chiral 3-point function} through the replacements $z_i \leftrightarrow \zbar_i$ and $h \leftrightarrow \bar h$ ($J_i \mapsto -J_i$).  
	
	\paragraph{4-point functions.}
	Momentum conservation with four momenta, when expressed in complex stereographic coordinates, amounts to \cite{Pasterski:2017ylz}
	\begin{equation}
	\label{z=zbar}
	z=\zbar\,,
	\end{equation}
	where $z, \zbar$ are the invariant cross ratios
	\begin{equation}
	z= \frac{z_{12}z_{34}}{z_{13}z_{24}}\,, \qquad \zbar= \frac{\zbar_{12}\zbar_{34}}{\zbar_{13}\zbar_{24}}\,.
	\end{equation}
	In the context of scattering amplitudes, we are thus interested in 4-point functions displaying a delta function $\delta(z-\zbar)$. We may also recall the useful relations
	\begin{equation}
	1-z=\frac{z_{14}z_{23}}{z_{13}z_{24}}\,, \qquad \frac{1-z}{z}=\frac{z_{14}z_{23}}{z_{12}z_{34}}\,.
	\end{equation}
	We again look for combinations of the coordinates $\x_i$ that are translation-invariant and transform covariantly under \eqref{complex coordinate transformations}. In addition to the separations $\x_{ij}$, on the support \eqref{z=zbar} we also have the combination
	\begin{equation}
	\label{F1234}
	\begin{split}
	F_{1234}&\equiv u_4-u_1 z \left| \frac{z_{24}}{z_{12}}\right|^2+u_2 \frac{1-z}{z}\left| \frac{z_{34}}{z_{23}}\right|^2-u_3 \frac{1}{1-z}\left| \frac{z_{14}}{z_{13}}\right|^2\\
	&=u_4-u_1\, \frac{z_{34} \zbar_{24}}{z_{13}\zbar_{12}}+u_2\, \frac{z_{14} \zbar_{34}}{z_{12}\zbar_{23}}-u_3\, \frac{z_{24}\zbar_{14}}{z_{23}\zbar_{13}}\,.
	\end{split}
	\end{equation}
	Note that, on the support \eqref{z=zbar}, any permutation on the indices yields a quantity related to \eqref{F1234} by a simple multiplicative factor, for instance
	\begin{equation}
	F_{4231}=-\frac{1}{z} \left|\frac{z_{12}}{z_{24}} \right|^2 F_{1234}\,, \qquad F_{2143}=-\frac{z_{13}\zbar_{23}}{z_{14}\zbar_{24}} F_{1234}\,, \qquad F_{1432}=\frac{z_{12}\zbar_{23}}{z_{14}\zbar_{34}} F_{1234}\,,
	\end{equation}
	where we note that under $1 \leftrightarrow 4$ we also have $z \leftrightarrow z^{-1}$.
	Therefore we can restrict our attention to $F_{1234}$ without loss of generality.
	\begin{mytheo}{}{}
		Show that $F_{1234}$ transforms as
		\begin{equation}
		\begin{aligned}
		F_{1234}'&=\lambda F_{1234}\,, &\qquad &(\text{dilation})\,,\\
		F_{1234}'&=\frac{F_{1234}}{1-k \zbar_4}\,, \quad F_{1234}'=\frac{F_{1234}}{1-\bar k z_4}\,, &\qquad &(\text{SCT})\,,
		\end{aligned}
		\end{equation}
		while, on the support $z=\zbar$, it is invariant under all other transformations \eqref{complex coordinate transformations}. 
	\end{mytheo}
	\noindent Thus we can assume an ansatz of the form
	\begin{equation}
	\label{4-point general}
	\langle O_1 O_2 O_3 O_4 \rangle= \delta(z-\zbar) G(z)\prod_{i<j} \frac{1}{(z_{ij})^{a_{ij}} (\zbar_{ij})^{\bar{a}_{ij}}(F_{1234})^c}\,,
	\end{equation}
	with $G(z)$ an arbitrary function of the invariant cross ratio, as required by translation and carroll boost invariance. 
	\begin{mytheo}{}{}
		Show that dilation, rotation, and special conformal transformations require    
		\begin{align}
		\label{aij}
		\begin{split}
		a_{ij}&=h_i+h_j-H/3+c/6\,, \quad (i,j \neq 4)\,,\\
		a_{i4}&=h_i+h_4-H/3-c/3\,,
		\end{split}
		\end{align}
		with $H\equiv \sum_i h_i$ and $c$ arbitrary, together with the conjugate relations. 
	\end{mytheo}
	\noindent Note that we are left with one free parameter $c$. If $c=0$ then \eqref{4-point general} reduces to a chiral four-point function familiar from conformal field theory on the complex plane, times the delta distribution $\delta(z-\zbar)$.\\
	
	\textit{Remark.} Actually, the formula \eqref{4-point general} is not the most general one. Indeed, one can form a second kinematic invariant on the support $z=\zbar$ given by
	\begin{equation}
	\mathcal{F}=F_{1234} \left|\frac{z_{23}}{z_{24}z_{34}} \right|\,.
	\end{equation}
	A general solution to the Ward identities is then given by
	\begin{equation}
	\langle O_1 O_2 O_3 O_4 \rangle= \delta(z-\zbar) G(z,\mathcal{F})\prod_{i<j} \frac{1}{(z_{ij})^{a_{ij}} (\zbar_{ij})^{\bar{a}_{ij}}}\,,
	\end{equation}
	where the exponents are still given simply given by $a_{ij}=h_i+h_j-H/3$ for all $i,j$. This more general form is needed in order to account for $\ln( \mathcal{F})$ terms arising in loop amplitudes \cite{Nenmeli:2026ket}.
	
	\subsection{Carrollian amplitudes}
	\label{subsec: carrollian amplitudes}
	We now turn to the computation of 2-, 3-, and 4-point carrollian amplitudes as defined in \eqref{Mellin amplitudes}, starting from well-known expressions of massless scattering amplitudes. This section mostly follows \cite{Mason:2023mti,Nguyen:2025sqk}. 
	
	\paragraph{2-point amplitudes.} This was already computed in \eqref{free transformed amplitude}.

	\paragraph{3-point amplitudes.} 
	As discussed at the beginning of this section, momentum conservation for three massless particles only leaves us with amplitudes that vanish as distributions if only real momenta are considered. With complex kinematics there exist 3-point amplitudes which are non-trivial even though they may appear unphysical. Their form is entirely fixed by the little group scalings and locality of the interaction, which is most conveniently displayed in spinor-helicity variables \cite{Elvang:2015rqa,Badger:2023eqz}
	\begin{align}
	\label{spinhel3pt}
	S_3(1^{J_1}2^{J_2}3^{J_3})=\begin{cases} \braket{12}^{J_3-J_1-J_2}\braket{31}^{J_2-J_1-J_3}\braket{23}^{J_1-J_2-J_3}\delta(\Sigma\, p), &J_1+J_2+J_3<0\,,\\
	[12]^{-J_3+J_1+J_2}[31]^{-J_2+J_1+J_3}[23]^{-J_1+J_2+J_3}\, \delta(\Sigma\, p), &J_1+J_2+J_3>0\,,\end{cases}
	\end{align}
	up to an overall free coefficient. As shown in \cite{Pasterski:2017ylz} the spinor-helicity variables can be chosen such that $\langle ij\rangle=\sqrt{\omega_i\omega_j}\,z_{ij}$ and $[ij]=-\eta_i\eta_j\sqrt{\omega_i\omega_j}\, \zb_{ij}$. The modified Mellin transform of \eqref{spinhel3pt} has been performed with $\Delta_k=1$ in \cite{Salzer:2023jqv,Mason:2023mti}. Generalizing their computation to arbitrary scaling dimensions yields \cite{Nguyen:2025sqk}, for $J_1+J_2+J_3<0$, 
	\begin{align}
	\label{eq:3ptwithTheta}
	\begin{split}
	\langle O_1 O_2 O_3 \rangle &=\Gamma[2\Sigma_k \bar h_k-4]\, \Theta\left(-\frac{z_{13}}{z_{23}}\eta_1\eta_2\right)\Theta\left(\frac{z_{12}}{z_{23}}\eta_1\eta_3\right)\\
	&\times \frac{\delta(\zbar_{12}) \delta(\zbar_{13}) (z_{12})^{\Delta_3-J_1-J_2-2} (z_{23})^{\Delta_1-J_2-J_3-2} (z_{13})^{\Delta_2-J_1-J_3-2}}{\left(z_{23}\, u_1-z_{13}\, u_2+z_{12}\, u_3 \right)^{2\Sigma_k \bar h_k-4}}\,,
	\end{split}
	\end{align}
	again up to an overall constant coefficient. We see that this is indeed of the general form \eqref{chiral 3-point function} derived previously on the basis of Poincar\'e covariance. The main additional feature is the presence of the Heaviside step functions. The expression for $J_1+J_2+J_3>0$ is obtained by the replacement $z_k \leftrightarrow \zbar_k$ and $h_k \leftrightarrow \bar h_k$.
	
	\paragraph{4-point amplitudes.}
	The simplest example of 4-particle scattering amplitude one can think of is the contact amplitude corresponding to $\lambda \phi^4$ interaction. The resulting 4-point carrollian amplitude is \cite{Nguyen:2025sqk}
	\begin{align}
	\begin{split}
	C_4&=\delta(z-\zbar)\Theta\left(-z\left|\frac{z_{24}}{z_{12}}\right|^2 \eta_1 \eta_4  \right)\Theta\left(\frac{1-z}{z} \left|\frac{z_{34}}{z_{23}}\right|^2\eta_2 \eta_4  \right)\Theta\left(-\frac{1}{1-z}\left|\frac{z_{14}}{z_{13}}\right|^2\eta_3 \eta_4 \right)\\
	&\times \frac{z^{\Delta_1-\Delta_2} (1-z)^{\Delta_2-\Delta_3}}{|z_{13}z_{24}|^2} \left|\frac{z_{24}}{z_{12}} \right|^{2(\Delta_1-1)}\left|\frac{z_{34}}{z_{23}} \right|^{2(\Delta_2-1)} \left|\frac{z_{14}}{z_{13}} \right|^{2(\Delta_3-1)}\frac{\Gamma[\Sigma \Delta-4]}{(iF_{1234})^{\Sigma \Delta-4}}\,.
	\end{split}
	\end{align}
	\begin{mytheo}{}{}
		Show that it can be written in the general form \eqref{4-point general} with 
		\begin{equation}
		G(z)=\left[z(1-z)\right]^{2/3}\,, \qquad c=\Sigma_i \Delta_i-4\,.
		\end{equation}
	\end{mytheo}
	\noindent This demonstrates the usefulness of \eqref{4-point general} in organizing 4-point carrollian amplitudes.
	
	Other examples could be given. For instance, one can look at maximal-helicity-violating (MHV) amplitudes of gluons and gravitons, which constitute an important class of amplitudes. The corresponding 4-point carrollian amplitudes may be found in \cite{Mason:2023mti,Nguyen:2025sqk}, which provide other realizations of the general formula \eqref{4-point general} allowed by kinematics.
	
	\section{Operator product expansion}
	\label{sec:OPE}
	In the previous sections we mostly discussed kinematic constraints on carrollian correlators, and it is now time to address the structure of interactions and the constraints they impose on the theory. It is time to tackle step 3 of the bootstrap program set out in section~\ref{subsec:bootstrap}.
	
	One of the pillars of conformal field theory is the operator product expansion (OPE), which allows to express the product of two local operators as a sum of local operators,
	\begin{equation}
	O_1(\vec x_1)\, O_2(\vec x_2)= \sum_{k} C_{12k}(\vec x_{12})\, O_k(\vec x_2) \,,
	\end{equation}
	where the sum is over primary operators and their descendants.  
	This equality is equivalent to the `state-operator correspondence' which expresses the fact that any quantum state can be created from insertion of a local operator at the point $\vec x_2$ \cite{Mack:1976pa}. It is usually presented as an expansion in the separation $\vec x_{12}$, such that it takes the form
	\begin{equation}
	\label{standard OPE coincident}
	O_1(\vec x)\, O_2(0)  \stackrel{\vec x \sim 0}{\approx } \sum_{k} \frac{c_{12k}}{|\vec x|^{\Delta_1+\Delta_2-\Delta_k}}\, O_k(0)+ subleading\,,
	\end{equation}
	where the subleading terms contain derivatives of the primary operators and therefore account for their descendants. The latter are actually completely fixed by conformal symmetry, such that the set of coefficients $\{c_{12k}\}$ carry all the independent data. 
	
	In this section we wish to investigate the existence of an analogous structure within carrollian conformal field theory. For simplicity we will first focus our attention on the weaker form of the OPE, i.e., presented as a series expansion in the variables $\x_{12}$. We will systematically construct the contributions from a given carrollian primary $O_k$ and its descendants, together forming a \textit{carrollian OPE block}. By contrast to standard CFT, a source of complexity comes from the fact that, as with correlation functions, the form of the leading term in the OPE is not completely fixed by symmetry. This leads to various possible OPE branches for a fixed $O_k$. Of course knowledge of the 3-point function $\langle O_1 O_2 O_k \rangle$ would determine the leading OPE coefficient and would thus select a particular OPE branch.
	
	We will then discuss the form of the carrollian OPE blocks for finite separation $\x_{12} \neq 0$, adapting the standard construction in \cite{Czech:2016xec}.

	\subsection{Short distance expansion}
	\label{section 5.1}
	In analogy with \eqref{standard OPE coincident}, we postulate the existence of an OPE of the form
	\begin{equation}
	\label{Carrollian OPE}
	O_1(\x)\, O_2(0) \stackrel{\x\sim 0}{\approx} \sum_k f_{12k}(\x)\, O_k(0)+subleading+massive\,,
	\end{equation}
	where the sum is over carrollian  primary fields, which we described in Section~\ref{sec: states and fields}. As usual the subleading terms involve the descendant operators, i.e., derivatives of the primary fields. It is a priori unclear whether such an OPE exists and what is the full set of operators which need to be considered on the right-hand side of \eqref{Carrollian OPE}. In particular, the `massive' terms may correspond to at least two different types of operators. First they may correspond to massive one-particle operators, for instance in the context of a scattering theory involving massive scattering states, in which case they cannot be local carrollian operators of the type considered here.\footnote{They are local carrollian operators at timelike infinity $\mathsf{Ti}$ \cite{Have:2024dff}.} Second they may correspond to multi-particle states. Although we will not consider the corresponding OPE blocks explicitly here, we will discuss their unavoidable appearance. Regardless, we proceed to constrain the functions $f_{12k}(\x)$ by requiring consistency with Poincar\'e symmetry. In practice we act on both sides of \eqref{Carrollian OPE} with the symmetry generators and require consistency order by order in $\x \sim 0$. 
	
	\subsubsection*{Several OPE branches} 
	We determine the explicit form of $f_{123}$ allowed by symmetry, focusing on the contribution from a single primary operator $O_3$. Acting with the Poincar\'e generators $\{ H,P_i\}$ does not yield any constraint since our ansatz already incorporates translation invariance on $\scri$. Acting with the generators $\{K,K_i,B_i\}$ on either side of \eqref{Carrollian OPE} does not contribute at leading order in $\vec x \sim 0$ as can be seen from \eqref{Carrollian induced rep}. Therefore we are left to act with $\{ D,J_{12} \}$ or equivalently with $\{L_0,\bar L_0\}$. Acting with $L_0$ on the left and on the right of \eqref{Carrollian OPE} yields, respectively,
	\begin{equation}
	\begin{split}
	\left[L_0,O_1(\x)\, O_2(0) \right]&=\left(\frac{u}{2}\partial_u+z\partial_z+h_1+h_2 \right)O_1(\x)\, O_2(0)\\
	&\approx \left(\frac{u}{2}\partial_u+z\partial_z+h_1+h_2 \right) f_{123}(\x)\, O_3(0)\,,
	\end{split}
	\end{equation}
	and
	\begin{equation}
	f_{123}(\x) \left[L_0,O_3(0) \right]=h_3\, f_{123}(\x)\, O_3(0)\,.
	\end{equation}
	Hence we should impose
	\begin{equation}
	\label{h definition}
	\left(\frac{u}{2}\partial_u+z\partial_z-h \right) f_{123}(\x)=0\,, \qquad h\equiv h_3-h_1-h_2\,,
	\end{equation}
	which essentially tells us that $f_{123}$ must have scaling weight $h=h_3-h_1-h_2$ under holomorphic scalings generated by $L_0$ (and similarly for $\bar L_0$). The general form satisfying this property is
	\begin{equation}
	\label{f12k}
	\begin{split}
	f_{123}(\x)&=c_0\, z^{h-a} z^{\bar h-a} u^{2a}+c_1\, \delta(z) \delta(\zbar)\, u^{h+\bar h+2}\\
	&+c_2\, \delta(\zbar) z^{h-\bar h-1} u^{2\bar h+2}+\bar c_2\, \delta(z) \zbar^{\,\bar h- h-1} u^{2h+2}\,,
	\end{split}
	\end{equation}
	where the coefficients $c_0,c_1,c_2,\bar c_2$ as well as the exponent $a$ are arbitrary numbers. The form of the OPE in carrollian CFT appears much less constrained than it is in conventional (two-dimensional) CFT, where $c_0$ would be the only nonzero parameter. 
	
	\subsubsection*{Chiral OPE branch}
	We start with the simple `chiral' OPE branch corresponding to $c_2\neq 0$ in \eqref{f12k}. We can start with the ansatz
	\begin{equation}
	\label{eq:deltaOPEansatz}
	O_1(\x)O_2(0)\sim \delta(\bar z) z^{h}\sum^\infty_{m,n=0}\alpha_{m,n}\frac{u^{m}}{m!}\frac{z^{n}}{n!} (P_{-1,-1})^m (L_{-1})^n\, O_3(0)\,,
	\end{equation}
	compatible with scale covariance.  Due to the the delta function, the only constraints on this ansatz come from covariance under $P_{0,-1}$ and $L_1$.
	\begin{mytheo}{}{}
		Act with $P_{0,-1}$ and $L_1$ on the left- and right-hand sides of \eqref{eq:deltaOPEansatz}, and show that the coefficients must be of the form
		\begin{equation}
		\label{eq:OPEcoefficientsdelta}
		\alpha_{m,n}
		=\alpha\, B(h_3+h_2-h_1,h_3-h_2+h_1+m+n)\,,
		\end{equation}
		with $B(x,y)$ the Euler beta function, and $\alpha$ some overall normalization.
	\end{mytheo}
	
	\subsubsection*{Other OPE branches}
	The chiral OPE branch was relatively straightforward to determine. The same is true of the `ultra-local' branch which corresponds to $c_1 \neq 0$ in \eqref{f12k}. The `regular' OPE branch, corresponding to $c_0 \neq 0$ in \eqref{f12k}, is significantly more involved. In standard conformal field theory, there is a finite number of operators which can appear at a given order of the short distance expansion. This is not the case anymore, since the operator $O_3$ of dimension $(h_3,\bar h_3)$ may possess \textit{parent primary operators} $O_{3'}$ of dimension $(h_{3'},\bar h_{3'})=(h_3-n/2,\bar h_3-n/2)$ in case they satisfy $O_3=(\partial_u)^n\, O_{3'}$. See \eqref{Mellin derivatives}. Moreover, the regular OPE branch requires to consider BMS descendants and not only Poincar\'e descendants \cite{Banerjee:2020kaa}. For a complete analysis, the interested reader should consult \cite{Nguyen:2025sqk}. In these lectures, we will content ourselves with the chiral OPE branch.

	\subsection{Finite distance resummation}
	\label{section 5.3}
	We now turn to the discussion of OPE blocks, first introduced in the context of standard conformal field theory in \cite{Czech:2016xec}. Their purpose is to resum the OPE \eqref{standard OPE coincident} such as to produce a formula valid for finite separation $\x_{12}\neq 0$. We adapt the discussion to the carrollian setup, assuming an ansatz of the form 
	\begin{equation}
	\label{carrollian OPE block}
	O_1(\x_1) O_2(\x_2)\sim \int_{\mathcal{D}(\x_1,\x_2)} d^3\x\, F_{12k}(\x_1,\x_2,\x)\, O_k(\x)\,,
	\end{equation}
	where $\mathcal{D}(\x_1,\x_2)$ is some domain of integration which depends on the operator insertions, and $F_{12k}(\x_1,\x_2,\x)$ is some kinematic function, both to be determined.
	Under coordinate transformations \eqref{complex coordinate transformations}, the integration measure transforms like
	\begin{equation}
	d^3\x'=\left(\frac{\partial z'}{\partial z} \right)^{3/2} \left(\frac{\partial \zbar'}{\partial \zbar} \right)^{3/2} d^3\x\,.
	\end{equation}
	Using the transformation law \eqref{conformal transfo} for the primary operator $O_k$, we have
	\begin{equation}
	\label{O1 O2 prime}
	\begin{split}
	O'_1(\x_1') O'_2(\x_2')&\sim \int_{\mathcal{D}(\x_1',\x_2')} d^3\x'\, F'_{12k}(\x_1',\x_2',\x')\, O'_k(\x')\\
	&=\int_{\mathcal{D}'(\x_1',\x_2')} d^3\x\, \left(\frac{\partial z'}{\partial z} \right)^{3/2-h_k} \left(\frac{\partial \zbar'}{\partial \zbar} \right)^{3/2-\bar h_k} F'_{12k}(\x_1',\x_2',\x')\, O_k(\x)\,.
	\end{split}
	\end{equation}
	Using the transformation of the operators $O_1(\x_1) O_2(\x_2)$, we must also have
	\begin{equation}
	\label{O1 O2 prime}
	\begin{split}
	O'_1(\x_1') O'_2(\x_2')&\sim \left(\frac{\partial z_1'}{\partial z_1} \right)^{-h_1}\left(\frac{\partial \zbar_1'}{\partial \zbar_1} \right)^{-\bar h_1} \left(\frac{\partial z_2'}{\partial z_2} \right)^{-h_2} \left(\frac{\partial \zbar_2'}{\partial \zbar_2} \right)^{-\bar h_2}\\
	&\times \int_{\mathcal{D}(\x_1,\x_2)} d^3\x\, F_{12k}(\x_1,\x_2,\x)\, O_k(\x)\,.
	\end{split}
	\end{equation}
	For consistency, $F_{12k}$ must therefore behave like a carrollian three-point function,
	\begin{equation}
	\label{eq:shadow3pt}
	F_{12k}(\x_1,\x_2,\x)=\langle O_1(\x_1) O_2(\x_2) \tilde O_k(\x) \rangle\,,
	\end{equation}
	where the \textit{shadow operator} $\tilde O_k$ has dimensions $\tilde h_k=3/2-h_k$ and $\tilde{\bar h}_k=3/2-\bar h_k$, or equivalently $\tilde \Delta=3-\Delta$ and $\tilde J=-J$. In addition, the domain of integration must be invariant under carrollian conformal transformations,
	\begin{equation}
	\mathcal{D}'(\x_1',\x_2')=\mathcal{D}(\x_1,\x_2)\,.
	\end{equation}
	For the spatial domain of integration, we can adopt the same one as in CFT$_2$, since carrollian conformal transformations \eqref{complex coordinate transformations} act as $2d$ conformal transformations on the celestial sphere. This is a diamond in the $(z,\zbar)$-plane, with edges given by $(z_1,\zbar_1)$ and $(z_2,\zbar_2)$ \cite{Czech:2016xec}. For the time domain, we could integrate $u$ over the whole real axis for instance, or define it as a closed contour in complex $u$-plane. We will leave it unspecified until needed.
	
	An important distinction compared to standard conformal field theory, is that there is a variety of three-point functions for any given set of fields \cite{Nguyen:2023miw,Nguyen:2025sqk}. Each allowed three-point function \eqref{eq:shadow3pt} defines an OPE block. These choices reproduce the various OPE branches found from \eqref{f12k} in the short distance limit \cite{Nguyen:2025sqk}.
	
	As an illustration, let us show that the chiral OPE branch \eqref{eq:deltaOPEansatz} is recovered from the OPE block formula \eqref{carrollian OPE block}, when the kinematic function \eqref{eq:shadow3pt} is chosen to be the 3-point carrollian amplitude \eqref{chiral 3-point function}. We thus write
	\begin{equation}
	\label{eq:chiralOPEblock}
	\begin{split}
	O_1(\x_1) O_2(\x_2)&=\int d^3\x_3 \frac{c_{123}\, \delta(\zbar_{12})\delta(\zbar_{23})}{ (z_{12})^a\, (z_{23})^b\, (z_{13})^c\, (F_{123})^d}\, O_3(\x_3)\\
	&=\frac{c_{123}\, \delta(\zbar_{12})}{z^{a+b+c+d-1}_{12}u^{d-1}_{12}}\int d t d s\, \frac{O_3(u_2+t u_{12},z_2+sz_{12},\zbar_2)}{(-s)^b(-1+s)^c(-s+t)^d}\,,
	\end{split}
	\end{equation}
	where in the second line we made the variable changes $u_3=u_2+tu_{12}$ and $z_3=z_2+s z_{12}$. The constants $a,b,c,d$ are given by \eqref{abcd} subject to the replacement $\Delta_3 \mapsto 3-\Delta_3$ and $J_3\mapsto -J_3$ -- the quantum numbers of the shadow operator. First, it can be easily checked that the leading term in the short distance expansion $u_{12}\,, z_{12}\sim 0$ agrees with that of \eqref{eq:deltaOPEansatz}.
	
	Consider now the special case $1-d=2\hb+2=0$ for which the leading $u$-dependence vanishes. We choose a contour in $u$, or equivalently $t$, that circles the pole in $(s-t)^{-1}$. Integrating over the domain $z_3\in(z_1,z_2)$, we thus have
	\begin{equation}
	\begin{split}
	O_1(\x_1) O_2(\x_2)&= -\frac{c_{123}\, \delta(\zbar_{12})}{z^{a+b+c}_{12}} \int^{1}_0 ds\, \frac{O_3(u_2+s u_{12},z_2+sz_{12},\zbar_2)}{s^b(1-s)^c}\\
	&=-\frac{c_{123}\, \delta(\zbar_{12})}{z^{a+b+c}_{12}}\int^{1}_0ds \sum^\infty_{m,n=0} \frac{u^m_{12}}{m!}\frac{z^n_{12}}{n!} s^{m+n-b}(1-s)^{-c}\, \partial^m_{u_2}\partial^n_{z_2}O_3(\x_2)\\
	&=-\frac{c_{123}\, \delta(\zbar_{12})}{z^{a+b+c}_{12}}\sum^\infty_{m,n=0}\frac{u^m_{12}}{m!}\frac{z^n_{12}}{n!} B(m+n+1-b,1-c)\, \partial^m_{u_2}\partial^n_{z_2}O_3(\x_2).
	\end{split}
	\end{equation}
	\begin{mytheo}{}{}
		Plug in the values of $a,b,c$ and show that this exactly agrees with \eqref{eq:deltaOPEansatz}-\eqref{eq:OPEcoefficientsdelta}.
	\end{mytheo}

	\subsection{Casimir constraint and two-particle operators} 
	\label{subsec:two particle operator}
	Let us now comment on the necessity of the `massive' terms for consistency of the carrollian operator product expansion \eqref{Carrollian OPE}. Without these we would be essentially proposing that a tensor product of two massless representations can be decomposed into massless representations, while it is well-known that massive representations also appear in general \cite{Barut:1986dd}. Put very simply, the total momentum of a pair of massless particles is not null,
	\begin{equation}
	\label{total momentum}
	(p_1+p_2)^2=2\, p_1 \cdot p_2 \propto \omega_1 \omega_2 |x_{12}|^2\,,    
	\end{equation}
	unless the particles are \textit{exactly} colinear or at least one of the momenta is zero.  
	In terms of carrollian operators, using the general identity
	\begin{equation}
	\left[AB,O_1O_2\right]=[A,[B,O_1]]O_2+O_1[A,[B,O_2]]+[A,O_1][B,O_2]+[B,O_1][A,O_2]\,,
	\end{equation} 
	we can evaluate the action of the quadratic Casimir operator $\mathcal{C}_2=-(HK+KH)+2B^i B_i$
	on the left-hand side of \eqref{Carrollian OPE}, yielding
	\begin{equation}
	\label{Casimir product}
	\begin{split}
	\left[\mathcal{C}_2, O_1 O_2\right]&=-2[H,O_1][K,O_2]-2[K,O_1][H,O_2]+4[B^i,O_1][B_i,O_2]\\
	&=2(x_1^2-2x_1 \cdot x_2+x_2^2)\partial_u O_1\partial_u O_2=2|x_{12}|^2\partial_u O_1\partial_u O_2\,.
	\end{split}
	\end{equation}
	This gives the total invariant mass of the product $O_1O_2$, which is indeed the same as \eqref{total momentum} modulo Fourier transform. 
	
	While \eqref{Casimir product} is generically nonzero, acting with  $\mathcal{C}_2$ on the right-hand side of \eqref{Carrollian OPE} would yield a strict zero if no massive operators were included, since $[\mathcal{C}_2,O]=0$ for any carrollian conformal primary field $O$. Thus `massive operators' need to be included in the carrollian OPE \eqref{carrollian OPE block}. One way to introduce massive operators that are still locally defined at $\scri$ is through indecomposable representations \cite{Nguyen:2025sqk}. The Casimir constraint \eqref{Casimir product} can then be accommodated order by order in $\x_{12}$, although more work is needed to find a resummation at finite $\x_{12}$.  
	
	\subsection{Basic example}
	
	In order to exemplify and check the relevance of the carrollian OPE studied in this section, we should investigate its realization within carrollian amplitudes. Here we look at one basic example, and refer the reader to \cite{Nguyen:2025sqk} for a more extensive study.   
	
	Let us consider the 3-point carrollian amplitude \eqref{eq:3ptwithTheta}, which we display again here for convenience,
	\begin{equation}
	\label{3-point amplitude OPE}
	\langle O_1 O_2 O_3 \rangle=\frac{\delta(\zbar_{12})\delta(\zbar_{23})}{ (z_{12})^a\, (z_{23})^b\, (z_{13})^c\, (F_{123})^d}\,,
	\end{equation}
	with $a,b,c,d$ given in \eqref{abcd}. We claim that it can be built from a \textit{single} chiral OPE block together with the 2-point amplitude. Thus, we set out to compute
	\begin{equation}
	\label{reconstruction formula}
	\langle O_1O_2O_3\rangle=\int d^3\x_4\, \frac{\delta(\zbar_{12})\delta(\zbar_{24})}{ (z_{12})^{\tilde a}\, (z_{24})^{\tilde b}\, (z_{14})^{\tilde c}\, (F_{124})^{\tilde d}}\, \langle O_4(\x_4) O_3(\x_3)\rangle\,,
	\end{equation}
	with $\tilde a,\tilde b,\tilde c,\tilde d$ given as in equations \eqref{abcd} upon replacing $\Delta_3 \mapsto 3-\Delta_4$ and $J_3 \mapsto -J_4$, with $(\Delta_4,J_4)$ the quantum numbers of the operator $O_4$. We also claim that the quantum numbers of the `exchanged' primary $O_4$ are
	\begin{equation}
	\label{Delta4 J4}
	\Delta_4=\Delta_1+\Delta_2-J_1-J_2-J_3-2\,, \qquad J_4=-J_3\,,\qquad \eta_4=-\eta_3.
	\end{equation}
	\begin{mytheo}{}{}
		Evaluate the parameters $\tilde a, \tilde b,\tilde c, \tilde d$, and obtain   
		\begin{equation}
		\begin{split}
		\tilde a&=\Delta_1+\Delta_2-J_3-3\,,\\
		\tilde b&=J_2+J_3-\Delta_1+2=b\,,\\
		\tilde c&=J_1+J_3-\Delta_2+2=c\,,\\
		\tilde d&=1\,.
		\end{split}
		\end{equation}
	\end{mytheo}
	\noindent Inserting the two-point carrollian amplitude
	\begin{equation}
	\langle O_4(\x_4) O_3(\x_3)\rangle=\frac{\delta(z_{34})\delta(\zbar_{34})}{(u_{34})^{\Delta_3+\Delta_4-2}}=\frac{\delta(z_{34})\delta(\zbar_{34})}{(u_{34})^{2(\hb_1+\hb_2+\hb_3-2)}}\,,
	\end{equation}
	we thus have
	\begin{equation}
	\langle O_1O_2O_3\rangle=\frac{\delta(\zbar_{12})\delta(\zbar_{23})}{ (z_{12})^{\tilde a}\, (z_{23})^{b}\, (z_{13})^{c}}\int   \frac{du_4}{(u_1 z_{23}+u_2 z_{31}+u_4 z_{12})\,(u_{34})^{2(\hb_1+\hb_2+\hb_3-2)}}\,.
	\end{equation}
	Now let us assume $2(\hb_1+\hb_2+\hb_3-2)=n+1$ with $n\in\mathbb{N}$ -- this condition is satisfied by the carrollian primaries \eqref{Mellin derivatives} -- such that we can integrate by parts and use the residue theorem,
	\begin{equation}
	\begin{split}
	\langle O_1O_2O_3\rangle&=\frac{\delta(\zbar_{12})\delta(\zbar_{23})}{ (z_{12})^{\tilde a}\, (z_{23})^{b}\, (z_{13})^{c}}\int   \frac{du_4}{(u_1 z_{23}+u_2 z_{31}+u_4 z_{12})\,(u_{34})^{n+1}}\\
	&=\frac{\delta(\zbar_{12})\delta(\zbar_{23})}{ (z_{12})^{\tilde a-n}\, (z_{23})^{b}\, (z_{13})^{c}} \int   \frac{du_4}{(u_1 z_{23}+u_2 z_{31}+u_4 z_{12})^{n+1}\,u_{34}}\\
	&= \frac{2\pi i\, \delta(\zbar_{12})\delta(\zbar_{23})}{ (z_{12})^{\tilde a-n}\, (z_{23})^{b}\, (z_{13})^{c}\, (F_{123})^{n+1}}\,. 
	\end{split}
	\end{equation}
	We note that $d=n+1$ and $\tilde a-n=a$, such that we have reconstructed \eqref{3-point amplitude OPE} from the contribution of a single chiral OPE block, as encapsulated by \eqref{reconstruction formula}.
	
\begin{figure}[h!]
		\centering
		\includegraphics[clip,scale=0.49]{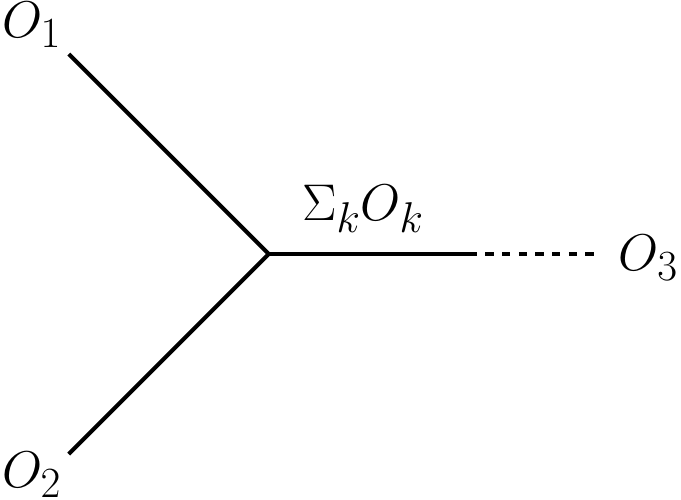}
		\hspace{0.4cm}
		\includegraphics[clip,scale=0.49]{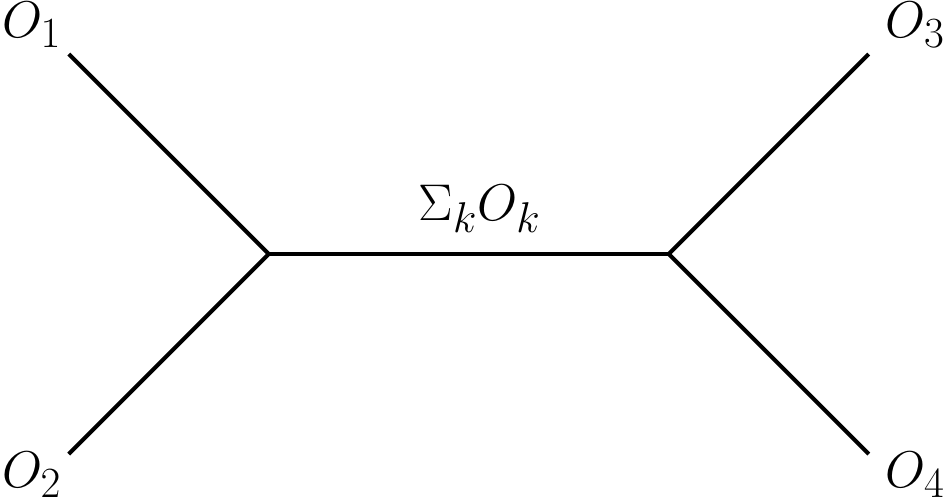}
		\caption{Application of the operator product expansion $O_1 O_2 \sim \Sigma_k O_k$ inside carrollian amplitudes. (Left) Only one OPE block contributes to the decomposition of a 3-point carrollian amplitude $\langle O_1 O_2 O_3 \rangle$, because $O_3$ projects out the all other blocks. (Right) Infinitely many OPE blocks are expected to contribute in the decomposition of 4-point carrollian amplitudes.}
		\label{OPE figure}
	\end{figure}

	In general, we expect that (infinitely) many OPE blocks contribute. The reason only one block contributes here is because the operator $O_3$ projects out all primary operators except $O_4$, with quantum numbers determined by \eqref{Delta4 J4}. See Figure~\ref{OPE figure}.
	
	Other examples worked out in \cite{Nguyen:2025sqk} give substantial evidence that the carrollian OPE controls the short-distance expansion of carrollian correlators and amplitudes. More work is needed however to complete step 3 of the bootstrap program set out in section~\ref{subsec:bootstrap}. In particular one needs to establish convergence of the carrollian OPE, including two-particle contributions as briefly discussed in section~\ref{subsec:two particle operator}. In other words, one needs to make sure that $(n+1)$-point amplitudes can be reconstructed from $n$-point amplitudes using the carrollian OPE. Achieving this would open up new ways to build and constrain carrollian amplitudes, and therefore scattering amplitudes, that are similar in spirit to the standard conformal bootstrap \cite{Poland:2018epd,Rychkov:2023wsd}.  
	
	\section{Soft theorems, symmetry breaking, and all that}
	\label{sec:soft theorems}
	In the previous sections, we have emphasized the development of a general framework to recast massless scattering amplitudes into carrollian conformal correlators on $\scri$, and we took the first steps towards `bootstrapping' these observables non-perturbatively. In this final section, we wish to show that Carrollian Holography also provides a new perspective on some well-known infrared physics, which will likely play a role in defining a consistent, non-perturbative gravitational scattering theory. This modern perspective on infrared gravitational physics goes back to seminal works of Strominger et al.~\cite{Strominger:2013jfa,He:2014laa,Strominger:2017zoo}, which predate and largely motivated the development of Celestial and Carrollian Holography.  
	
	\subsection{Soft theorems and infrared divergences}
	The year 1965 marks the publication of Weinberg's landmark paper on ``Infrared photons and gravitons'' \cite{Weinberg:1965nx}, where he provided a transparent derivation of the \textit{soft photon and soft graviton theorems}, together with the infrared divergences they imply at one-loop and beyond.   
	
	The soft graviton theorem relates the amplitude for scattering $n$ arbitrary particles plus one graviton of small energy, with the amplitude for scattering only the same $n$ particles, namely
	\begin{equation}
	\label{soft graviton theorem}
	\lim_{\omega \to 0} \omega\, S_{n+1}\left(p_1\,,...\,, p_n\,; \omega q,\varepsilon_{\mu\nu}(q)\right)=\kappa \sum_{a=1}^n \frac{\varepsilon_{\mu\nu}(q) p_a^\mu p_a^\nu}{p_a \cdot q}\, S_n\left(p_1\,,...\,, p_n\right)\,,
	\end{equation}
	where $\omega q$ is the null momentum of the graviton, $\varepsilon_{\mu\nu}(q)$ its polarization tensor, and $\kappa=\sqrt{8\pi G}$. The interest of this soft graviton theorem is that it is universal to all (perturbative) theories of gravity in four dimensions, and that it is valid to any given loop order. The only assumption needed is the coupling $\kappa\, h_{\mu\nu} T^{\mu\nu}$ between the graviton field and the rest of matter. For these reasons, soft theorems have become an integral part of modern on-shell approaches to scattering amplitudes \cite{Elvang:2015rqa}. 
	
	The most basic feature of the soft graviton theorem \eqref{soft graviton theorem} is that it predicts a $\omega^{-1}$ divergence at small frequency $\omega\ll 1$. Using this, Weinberg was able to argue that computing $S_n$ to the next loop order yields an infrared divergence (in four dimensions). Iterating this process to arbitrary loop order, the resulting infrared divergences can be re-summed into an exponential factor, 
	\begin{equation}
	\label{Sn all loop}
	S_n^{(all-loop)}=\exp\left[\frac{1}{\epsilon} \frac{\kappa^2}{(8\pi)^2} \sum_{a,b} (p_a \cdot p_b) \ln \left(\frac{p_a\cdot p_b}{\mu^2}\right) \right] S_n^{(finite)}\,,
	\end{equation}
	where $\epsilon>0$ is an infrared regulator. The energy scale $\mu$ has been introduced for dimensional reasons but is irrelevant on account of momentum conservation $\Sigma_a\, p_a=0$. The exponential prefactor re-sums the infrared divergences arising from all loops, while $S_n^{(finite)}$ is finite in the limit $\epsilon \to 0^+$. This result is highly problematic. Indeed, as the infrared regulator is removed, the exponent tends to minus infinity and the all-loop amplitude simply vanishes! This seems to imply that there is \textit{no gravitational S-matrix} ... These arguments apply equally well to QED. The commonly adopted resolution to this problem is to consider `inclusive cross-sections' \cite{Bloch:1937pw}, where one sums over diagrams with arbitrary number of soft photon/graviton emissions, and integrates over the soft momenta up to some physical detector threshold. Doing these phase space integrals produces additional infrared divergences which precisely cancel those encapsulated by the exponential prefactor in \eqref{Sn all loop}, giving a finite inclusive cross section as $\epsilon \to 0^+$ \cite{Weinberg:1965nx}. The physical interpretation given to this result is that classical soft radiation is always produced in any electromagnetic or gravitational scattering experiment. In other words, the probability to scatter a finite number of quanta is strictly zero. 
	
	Although this is a valid physical picture, from a theoretical and computational perspective, this state of affairs is quite unsatisfactory. In practice, it is cumbersome to compute the finite inclusive cross sections (see \cite{Agarwal:2021ais} for an introductory review), and it would be better to formulate a theory where physical observables can be computed without the need to introduce infrared regulators. If we wish to formulate a non-perturbative gravitational scattering theory, we will eventually have to face this issue.
	
	\subsection{Spontaneous breaking of asymptotic symmetries}
	Asymptotic symmetries have offered new insights about soft theorems and infrared divergences. Because BMS symmetries are most easily thought of conformal isometries of $\scri$, the carrollian framework that we have described in the previous sections is particularly well suited to present these new developments. 
	
	Adopting the framework of Carrollian Holography, we will observe that Weinberg's soft graviton theorem \eqref{soft graviton theorem} can be regarded as a Ward identity for \textit{spontaneously broken supertranslations}. The formulation proposed here is taken from \cite{Agrawal:2025bsy}, where an analogous treatment is also given for the soft photon theorem.
	
	We thus return to the BMS symmetries introduced in Section~\ref{subsec:BMS}, viewed as (globally defined) conformal isometries of $\scri$. The corresponding group is a semidirect product,
	\begin{equation}
	\label{BMS group}
	\operatorname{BMS}=\operatorname{SO}(1,3) \ltimes \mathcal{E}_{-1}(S^2)\,.
	\end{equation}
	An element of the supertranslation abelian subgroup $\mathcal{E}_{-1}(S^2)$ is a smooth function $T(\vec x)$ over the celestial sphere, which we again cover with stereographic coordinates $\vec x$. The Lorentz group $\operatorname{SO}(1,3)$ acts as the group of conformal isometries of the sphere $S^2$, and its action on the function $T(\vec x)$ is specified by saying that the latter transforms as a conformal primary field of scaling dimension $\Delta_T=-1$. The action of the supertranslation generators on the carrollian conformal fields~\eqref{Mellin} are simply obtained from \eqref{carrollian tensor} and $\xi=f(\vec x)\partial_u$. Denoting by $Q[f]$ the corresponding generator, this is simply given by 
	\begin{equation}
	[Q[f]\,, O_{\Delta,J}(\x)]=i  f(\vec x)\, \partial_u O_{\Delta,J}(\x)\,.
	\end{equation}
	This essentially means that supertranslations couple to the energy of the massless particles.
	
	As usual, symmetries yield Ward identities satisfied by correlation functions. In particular, computing the expectation value of $[Q[f]\,, O_1(\x_1)\,...\,O_n(\x_n)]$ in the vacuum state $|0 \rangle$ yields the Ward identity
	\begin{equation}
	\label{graviton Ward identity}
	\begin{split}
	&\langle 0|Q[f]\,  O_1(\x_1)\,...\,O_n(\x_n) - O_1(\x_1)\,...\,O_n(\x_n)\, Q[f] |0\rangle\\
	&= i\sum_{a=1}^n  f(\vec x_a)\partial_{u_a} \langle 0| O_1(\x_1)\,...\,O_n(\x_n) |0\rangle\,.
	\end{split}
	\end{equation}
	These equations can be used to \textit{diagnose} spontaneous breaking of asymptotic symmetries. Indeed, if they are unbroken, then $Q[f]|0\rangle=0$ and the left-hand side of \eqref{graviton Ward identity} vanishes. Conversely, evaluating the right-hand side and finding a nonzero result implies spontaneous symmetry breaking, $Q[f]|0\rangle\neq 0$. Weinberg's soft graviton theorem will select this second option.  
	
	To see this, we consider all particles to be massless and we adopt the momentum parametrization \eqref{massless momentum parametrisation}. We thus have
	\begin{equation}
	\lim_{\omega \to 0} \omega\, S_{n+1}\left(...; \omega q(\vec y),\varepsilon_{\mu\nu}(\vec y)\right)=-\kappa \sum_{a=1}^n \eta_a\omega_a\,  \frac{\varepsilon_{\mu\nu}(\vec y) q^\mu(\vec x_a) q^\nu(\vec x_a)}{|\vec y-\vec x_a|^2}\, S_n\left(...\right)\,.
	\end{equation}
	Applying the Fourier-Mellin transform \eqref{Mellin} to the $n$ `hard' particles, we can write\footnote{We use the S-matrix identity \cite{He:2014cra}
		\begin{equation*}
		\lim_{\omega \to 0} \omega\, \langle \text{out}| \text{S} a^\dagger(\omega q)|\text{in} \rangle=-\lim_{\omega \to 0} \omega \langle \text{out}|a(\omega q) \text{S} |\text{in} \rangle\,.
		\end{equation*}}
	\begin{equation}
	\label{carrollian soft graviton theorem}
	\begin{split}
	\lim_{\omega \to 0} \omega \left[\langle O_1(\x_1)...O_n(\x_n) a^\dagger_i(\omega q(\vec y)) \rangle- \langle a_i(\omega q(\vec y)) O_1(\x_1)...O_n(\x_n) \rangle \right]\\ 
	=\frac{i\kappa}{2} \sum_{a=1}^n \frac{\varepsilon_{\mu\nu}(\vec y) q^\mu(\vec x_a) q^\nu(\vec x_a)}{|\vec y-\vec x_a|^2}\, \partial_{u_a} \langle O_1(\x_1)...O_n(\x_n) \rangle\,.
	\end{split}
	\end{equation}
	The right-hand side of this equation coincides with the right-hand side of \eqref{graviton Ward identity} provided we choose the particular symmetry parameter
	\begin{equation}
	f(\vec x)=\frac{\kappa}{2}\frac{\varepsilon_{\mu\nu}(\vec y) q^\mu(\vec x) q^\nu(\vec x)}{|\vec y-\vec x|^2}\,,
	\end{equation}
	determined in terms of the momentum coordinate $\vec y$ and the polarization tensor of the soft graviton. Since this right-hand side is non-zero, it implies spontaneous breaking of BMS supertranslations. To complete the correspondence between \eqref{graviton Ward identity} and \eqref{carrollian soft graviton theorem}, we further identify a soft graviton state with the shifted vacuum,
	\begin{equation}
	\label{shifted vacuum}
	Q[f] |0\rangle \sim \lim_{\omega \to 0} \omega a^\dagger (\omega q(\vec y))|0 \rangle\,.
	\end{equation}
	
	We have thus established, using the carrollian description of amplitudes, that Weinberg's soft theorem implies the spontaneous breaking pattern
	\begin{equation}
	\operatorname{BMS} \quad  \stackrel{\text{SSB}}{\longrightarrow} \qquad  \operatorname{ISO}(1,3)\,.
	\end{equation}
	Note that translations are unbroken. Indeed in that case the right-hand side of \eqref{graviton Ward identity} do evaluates to zero on account of momentum conservation.  
	
	\subsection{Celestial Goldstone particles}
	Let us study the implications of spontaneous symmetry breaking on the spectrum of the theory. Within relativistic quantum field theory, spontaneous symmetry breaking typically implies the existence of massless particles in the spectrum \cite{Burgess:2020tbq}. In a carrollian conformal field theory, we will show that spontaneous symmetry breaking implies the existence of \textit{zero-momentum} representations in the spectrum.
	
	First we recall the explicit formula \eqref{BMS charges} for the supertranslation charges,
	\begin{equation}
	\label{integral charges}
	Q[f](u) =\oint_{\mathbb{S}_u} \dd^2 \vec x\, f(\vec x)\, M(u,\vec x)\,,
	\end{equation}
	written in terms of the mass aspect $M(\x)$. The charges \eqref{integral charges} are true canonical generators of supertranslations only when $\mathbb{S}_u$ lies close to spatial infinity, i.e., when $u \to -\infty$. This is the region of spacetime where radiation is absent, and where the mass aspect is actually conserved. 
	
	The existence of a spontaneously broken symmetry can be characterized by the existence of a field $O$ that is charged under this symmetry, with symmetry transformation given by
	\begin{equation}
	\label{delta psi}
	[ Q[f]\,, O]=\phi[f]\,,
	\end{equation}
	such that $\phi[f]$ has nonzero expectation value in vacuum, 
	\begin{equation}
	\label{order parameter}
	\langle 0|\phi[f]|0\rangle \neq 0\,.
	\end{equation}
	This quantity, usually called order parameter, detects spontaneous symmetry breaking. Indeed, this is only possible if the vacuum breaks the symmetry,
	\begin{equation}
	Q[f]|0\rangle \neq 0\,.
	\end{equation}
	In the previous section we had considered the charged operator $O=O_1(\x_1)\,...\,O_n(\x_n)$ to detect this spontaneous symmetry breaking, see \eqref{graviton Ward identity}.
	We can then write, 
	\begin{equation}
	\begin{split}
	\langle 0|\phi[f]|0\rangle&=\langle 0|[Q[f]\,, O]|0\rangle=\int d^2\vec x\, f(\vec x) \langle 0| [M(\x)\,,O]|0\rangle \big|_{u\to -\infty}\\
	&=\sum_n \int d^2\vec x\, f(\vec x) \left[ \langle 0| M(\x)|n\rangle \langle n|O|0\rangle -\langle 0| O|n\rangle \langle n|M(\x)|0\rangle\right]\big|_{u\to -\infty}\,,
	\end{split}
	\end{equation}
	where in the second line we have inserted a resolution of the identity on the Hilbert space of the theory, schematically denoted $1=\Sigma_n |n\rangle \langle n|$. Since this quantity is nonzero by \eqref{order parameter}, there must exist at least one particle species $|G\rangle$ such that 
	\begin{equation}
	\label{rho G}
	\langle 0|M(\x)|G\rangle\big|_{u\to -\infty} \neq 0\,.
	\end{equation}
	We wish to determine the properties of this type of Goldstone particle.
	
	First, let us use the retarded time translation operators $H$ to write 
	\begin{equation}
	M(\x)=e^{-iuH}  M(0,\vec x)\, e^{iuH}\,,
	\end{equation}
	such that 
	\begin{equation}
	\langle 0|M(\x)|G\rangle =\langle 0|M(0,\vec x)\, e^{iuH}|G\rangle\,,
	\end{equation}
	where we have assumed the vacuum to be time-translation invariant, i.e., $H|0\rangle=0$. 
	Then the Bondi mass loss formula \eqref{Bondi mass loss} near $u \to -\infty$ provides the appropriate `carrollian current conservation', and in particular implies
	\begin{equation}
	\label{current conservation matrix element}
	0=\langle 0| \partial_u M(\x)|G\rangle=\partial_u \langle 0|M(0,\vec x)\, e^{iuH}|G\rangle \big|_{u\to -\infty}\,.
	\end{equation}
	This can only be satisfied if the Goldstone particle is also invariant under retarded time translation,
	\begin{equation}
	H |G\rangle=0\,.
	\end{equation}
	This can only happens if the Goldstone particle belongs to a \textit{zero-momentum} representation, $\tilde P_\mu |G\rangle=0$, i.e., a simple unitary irrep of $\operatorname{SO}(1,3)$. We invite the reader to consult \cite{Sun:2021thf} for a friendly review of UIRs of $\operatorname{SO}(1,3)$. The upshot is that they can all be given in terms of conformal primary states $|\vec x\rangle_{\Delta,J}$, transforming under Lorentz according to 
	\begin{align}
	P_i\, |\vec x\rangle_{\Delta,J} &=-i \partial_i\, |\vec x\rangle_{\Delta,J}\,,\nonumber\\
	\label{induced rep}
	J_{ij}\, |\vec x\rangle_{\Delta,J}&=-i\left(x_i \partial_j-x_j\partial_i+i J \varepsilon_{ij} \right) |\vec x\rangle_{\Delta,J}\,,\\
	D\, |\vec x\rangle_{\Delta,J}&=i\left(\Delta+x^i\partial_i \right)|\vec x\rangle_{\Delta,J}\,,\nonumber\\
	K_i\, |\vec x\rangle_{\Delta,J}&=i\left(2x_i \Delta+2x_i x^j \partial_j-x^2 \partial_i+2i J x^j \varepsilon_{ij} \right)|\vec x\rangle_{\Delta,J}\,.\nonumber
	\end{align}
	Hence a zero-momentum state corresponds to a time-independent carrollian conformal field, or more simply to a $\operatorname{SO}(1,3)$ conformal field on the celestial sphere,
	\begin{equation}
	\label{Goldstone state}
	G_{\Delta,J}(\vec x)|0\rangle=|\vec x \rangle_{\Delta,J}\,.
	\end{equation}
	This is indeed in perfect agreement with early literature where Goldstone modes have been introduced as conformal fields on the celestial sphere \cite{Nande:2017dba,Himwich:2020rro}. We are left to determine the quantum numbers $(\Delta,J)$. Following the discussion around \eqref{BMS group}, and the fact that Goldstone modes look like `dynamical symmetry parameters', the quantum numbers of the supertranslation Goldstone is identified to be $\Delta=-1,J=0$. We note that this corresponds to a UIR in the discrete unitary series (the shadow dimension $\Delta_s=2-\Delta$ is a positive integer). 
	
	Importantly, the inner product on the Hilbert space corresponding to the unitary discrete series with $\Delta_s=2-\Delta \in \mathbb{N}$ is determined, up to some normalization constant $\mathcal{N}$, by the kernel \cite{Dobrev:1977qv,Sun:2021thf}\footnote{The inner product for states of the form
		\begin{equation}
		|\psi \rangle=\int d^2\vec x\, \psi(\vec x) |\vec x\rangle\,,
		\end{equation}
		does not actually depend on the arbitray scale $\mu$ as explained in \cite{Dobrev:1977qv,Sun:2021thf}.}
	\begin{equation}
	\label{kernel}
	\langle \vec x_1|\vec x_2 \rangle=\mathcal{N}\, |\vec x_{12}|^{-2\Delta} \ln (\mu |\vec x_{12}|)\,,
	\end{equation}
	which is also the two-point function of the Goldstone field according to \eqref{Goldstone state}. Here we see that such a logarithmic two-point function arises naturally and unavoidably from the nature of the Hilbert space. 
	
	\subsection{Infrared divergences through the looking-glass}
	Having established the existence of a new kind of Goldstone `particle', let us have another look at the re-summed infrared divergences in \eqref{Sn all loop}. Again assuming all particles to be massless and adopting the parametrization \eqref{massless momentum parametrisation}, it reads
	\begin{equation}
	S_n^{(all-loop)}=\exp\left[- \frac{2\kappa^2}{\epsilon (8\pi)^2} \sum_{a,b} \eta_a \eta_b\, |\vec x_{ab}|^2 \ln (\mu|\vec x_{ab}|)\right] S_n^{(finite)}\,,
	\end{equation}
	What appears in the exponent is precisely the two-point function \eqref{kernel} of the supertranslation Goldstone particle! This was proposed already a few years ago \cite{Nande:2017dba,Himwich:2020rro}, but the interpretation in terms of a \textit{unitary} irrep of the Lorentz group is fairly new \cite{Agrawal:2025bsy}. 
	
	Furthermore, let us add that the $\omega^{-1}$ behavior implied by the soft theorem \eqref{soft graviton theorem} is simply \textit{inconsistent} with the Hilbert space of photons and gravitons, as discussed for example in \cite{Ashtekar:2018lor,Satishchandran:2019pyc}. The norm of a photon/graviton state $|\psi \rangle$ of the form
	\begin{equation}
	|\psi \rangle =\sum_{i=+,-} \int d^2\vec x\int_0^\infty d\omega\, \psi_i(\omega,\vec x) |p(\omega,\vec x)\rangle_i\,,
	\end{equation}
	is given by
	\begin{equation}
	|| \psi ||^2=\sum_{i=+,-}\int d^2\vec x\, \int_0^\infty d\omega\, \omega\, |\psi_i(\omega,\vec x)|^2\,,
	\end{equation}
	and a wavefunction $\psi_i(\omega) \sim \omega^{-1}$ yields a logarithmic divergence from the lower bound of the integral. What the picture given here suggests is to associate Weinberg's soft pole not with conventional photon or graviton states of vanishing energy, but rather with a shifted vacuum state \eqref{shifted vacuum}. The corresponding Goldstone excitation comes equipped with its own Hilbert space, and seems to be in control of infrared divergences \cite{Nande:2017dba,Himwich:2020rro,Nguyen:2021ydb,Kapec:2021eug,Donnay:2022hkf,Agrawal:2023zea,Nguyen:2023ibj}. Hopefully, this new perspective can help us properly define a non-perturbative gravitational scattering theory, but this is still yet to come! 
	
	\section{Further reading}
	
	Here is some suggestion for further reading on various aspects of Carrollian Holography and closely related topics.
	
	\paragraph{Carrollian correlators and amplitudes.}  There is already a relatively large body of literature on carrollian amplitudes and their relation to massless scattering amplitudes
	\cite{Banerjee:2018gce,Banerjee:2019prz,Banerjee:2020kaa,Donnay:2022wvx,Bagchi:2022emh,Saha:2023hsl,Salzer:2023jqv,Nguyen:2023vfz,Saha:2023abr,Nguyen:2023miw,Bagchi:2023cen,Mason:2023mti,Chen:2023naw,Stieberger:2024shv,Kraus:2024gso,Kraus:2025wgi,Banerjee:2024hvb,Ruzziconi:2024zkr,Ruzziconi:2024kzo,Liu:2022mne,Liu:2024llk,Liu:2024nfc,Nguyen:2025sqk,Kulkarni:2025qcx}. The theoretical foundations for incorporating \textit{massive} amplitudes within Carrollian Holography have been described in \cite{Have:2024dff}, which involves the inclusion of timelike infinity as part of the boundary description.
	
	\paragraph{Carrollian limit of AdS/CFT.} The philosophy behind Carrollian Holography largely borrows from the very powerful AdS/CFT correspondence \cite{Aharony:1999ti}. The possibility to learn about Carrollian Holography from a suitable ultra-relativistic/carrollian limit of AdS/CFT is the subject of ongoing investigations \cite{Bagchi:2023fbj,Campoleoni:2023fug,Alday:2024yyj,Kraus:2024gso,Lipstein:2025jfj,Fontanella:2025tbs,Surubaru:2025fmg,Hao:2025btl,deGioia:2025mwt,Berenstein:2025tts}.
	
	\paragraph{Carroll field theories.}  Substantial effort is also being put in constructing explicit models of (non-conformal) carroll field theories and in investigating their peculiar features \cite{Bagchi:2016geg,Bagchi:2019xfx,Bagchi:2019clu,Gupta:2020dtl,Henneaux:2021yzg,deBoer:2021jej,Chen:2021xkw,Chen:2023pqf,Chen:2024voz,Baiguera:2022lsw,Rivera-Betancour:2022lkc,deBoer:2023fnj,Ciambelli:2023xqk,Bagchi:2024unl,Cotler:2024xhb,Cotler:2025dau,Poulias:2025eck,Despontin:2025dog,Aggarwal:2025hji}. See also the recent review \cite{Bagchi:2025vri}. 
	
	\paragraph{Carroll geometry and hydrodynamics.} Hydrodynamics is a useful framework to deal with dynamics at a macroscopic level. It is closely related to the study of curved geometry, as the stress tensor and other universal quantities can be viewed as geometric response functions. Studies of carrollian geometry and of the corresponding hydrodynamics can be found in   \cite{Ciambelli:2018xat,Ciambelli:2018ojf,Petkou:2022bmz,Freidel:2022bai,Mittal:2022ywl,Armas:2023dcz,Bagchi:2023ysc,Bagchi:2023rwd,Kolekar:2024cfg,Ciambelli:2025unn}.
	
	\paragraph{Fluid/gravity correspondence.} There is a clear correspondence between Einstein equations evaluated near null infinity $\scri$ and the hydro equations for carrollian fluids. See \cite{Ciambelli:2018wre,Ciambelli:2019lap,Ciambelli:2020ftk,Campoleoni:2022wmf,Hartong:2025jpp,Fiorucci:2025twa,Arenas-Henriquez:2025rpt}.
	
	\paragraph{Asymptotic symmetries.} The study of BMS symmetries \cite{Bondi:1962px,Sachs:1962wk,Sachs:1962zza} and extensions thereof \cite{Barnich:2009se,Barnich:2010eb,Campiglia:2014yka,Campiglia:2015yka} has a long history -- see the long list of references in \cite{Nguyen:2022nnx}. One important aspect to mention is that they are realized in all asymptotic regions of spacetime: null infinity $\scri^\pm$ \cite{Geroch1977,Barnich:2010eb,Barnich:2009se,Barnich:2011mi,Barnich:2016lyg,Compere:2018ylh,Geiller:2022vto,Geiller:2024ryw}, spatial infinity $i^0$ \cite{Ashtekar:1978zz,Ashtekar:1990gc,Ashtekar:1991vb,Compere:2011ve,Troessaert:2017jcm,Henneaux:2018cst,Henneaux:2019yax,Fiorucci:2024ndw}, and timelike infinity $i^\pm$ \cite{Campiglia:2015kxa,Chakraborty:2021sbc,Compere:2023qoa,Have:2024dff}. Moreover, these seemingly distinct realizations are connected to one another \cite{Troessaert:2017jcm,Capone:2022gme,Compere:2023qoa}. Recent work on BMS symmetries that are more directly connected to Carrollian Holography include \cite{Bagchi:2010zz,Barnich:2021dta,Donnay:2021wrk,Figueroa-OFarrill:2021sxz,Barnich:2022bni,Nguyen:2020hot,Donnay:2022aba,Donnay:2024qwq,Baulieu:2025itt}. 
	
	\paragraph{Soft theorems.} Soft theorems provide physical implications of asymptotic symmetries at the level of scattering amplitudes. It has been established over the last decade that \textit{leading} soft theorems \cite{Weinberg:1965nx} are consequences of Ward identities associated with asymptotic symmetries \cite{Strominger:2013lka,Strominger:2013jfa,He:2014cra,Campiglia:2015qka,Kapec:2015ena,Strominger:2017zoo}. This was the subject of section~\ref{sec:soft theorems}. Further \textit{subleading} soft theorems have been formulated \cite{Low:1958sn,Cachazo:2014fwa,Elvang:2016qvq,Laddha:2017vfh,Laddha:2018myi,Sahoo:2018lxl}, and shown equivalent to asymptotic conservation laws \cite{Lysov:2014csa,Adamo:2014yya,Kapec:2014opa,Campiglia:2015kxa,Conde:2016csj,Campiglia:2018dyi,Campiglia:2019wxe,AtulBhatkar:2019vcb,Donnay:2022hkf,Agrawal:2023zea,Choi:2024mac,Choi:2024ajz,Compere:2025tzr}. 
	
	\paragraph{Celestial holography.} The literature on Celestial Holography is very extensive. We recommend the reader to consult the reviews \cite{Pasterski:2021raf,McLoughlin:2022ljp,Donnay:2023mrd}. See also \cite{Donnay:2022wvx,Donnay:2022aba} for a direct connection to Carrollian Holography.
	
	\section*{Acknowledgments}
	I thank St\'ephane Detournay, Frank Ferrari and Marc Henneaux for encouraging me giving these lectures, and the International Solvay Institutes for their logistic support. I also thank Shreyansh Agrawal, Glenn Barnich, Laura Donnay, Emil Have, Lorenzo Iacobacci, Stefan Prohazka, Romain Ruzziconi, Jakob Salzer, Giovanni Spinielli, and Peter West for past collaboration on topics related to carrollian holography. This work is supported by a Postdoctoral Research Fellowship granted by the F.R.S.-FNRS (Belgium).
	
	\bibliography{bibl}
\end{document}